%% file: main.tex
\documentclass[journal]{IEEEtran}
\usepackage{cite}

\usepackage{subcaption}
\usepackage{amsmath}
\usepackage{amsthm}
\usepackage{amssymb}
\usepackage{array}
\usepackage{url}
\usepackage{graphicx}
\usepackage{lipsum}
\usepackage{xcolor}
\usepackage{multirow}
\usepackage{nomencl}
\usepackage{array}

\usepackage[acronym]{glossaries}

\makeglossaries

\newacronym{phy}{PHY-security}{Physical Layer Security}
\newacronym{ota}{OTA}{Over the Air}
\newacronym{afc}{AFC}{Analog Function Computation}
\newacronym{ss}{SS}{Spectrum Sensing}
\newacronym{sc}{SC}{Superposition Coding}
\newacronym{cdma}{CDMA}{Code Division Multiple Access}
\newacronym{mimo}{MIMO}{Multiple Input Multiple Output}
\newacronym{noma}{NOMA}{Non Orthogonal Multiple Access}
\newacronym{sic}{SIC}{Successive Interference Cancellation}
\newacronym{ml}{ML}{Machine Learning}
\newacronym{dsgd}{DSGD}{Distributed Stochastic Gradient Descent}
\newacronym{ofdm}{OFDM}{Orthogonal Frequency Division Multiplexing}
\newacronym{wmac}{W-MAC}{Wireless Multiple Access Channel}
\newacronym{mac}{MAC}{Multiple Access Channel}
\newacronym{csi}{CSI}{Channel State Information}
\newacronym{sdr}{SDR}{Software Defined Radio}
\newacronym{fc}{FC}{Fusion Fenter}
\newacronym{cfo}{CFO}{Carrier Frequency Offset}
\newacronym{mse}{MSE}{Mean Square Error}
\newacronym{mmse}{MMSE}{Minimum Mean Square Error}
\newacronym{stac}{STAC}{Simultaneous Transmitting and Air Computing}
\newacronym{iot}{IoT}{Internet of Things}
\newacronym{uav}{UAV}{Unmanned Aerial Vehicles}
\newacronym{llr}{LLR}{Log-Likelihood Ratio}
\newacronym{tbma}{TBMA}{Type-based Multiple Access}
\newacronym{tdma}{TDMA}{Time Division Multiple Access}
\newacronym{iid}{i.i.d.}{independent and identically distributed}
\newacronym{awgn}{AWGN}{Additive White Gaussian Noise}
\newacronym{wsn}{WSN}{Wireless Sensor Network}
\newacronym{snr}{SNR}{Signal to Noise Ratio}
\newacronym{plnc}{PLNC}{Physical Layer Network Coding}
\newacronym{nc}{NC}{Network Coding}
\newacronym{ldpc}{LDPC}{Low Density Parity Check}
\newacronym{ber}{BER}{Bit Error Rate}
\newacronym{ser}{SER}{Symbol Error Rate}
\newacronym{cee}{CEE}{Channel Estimation Error}
\newacronym{inc}{INC}{In-Network Computation}
\newacronym{map}{MAP}{Maximum A Posteriori}
\newacronym{simo}{SIMO}{Single Input Multiple Output}
\newacronym{cfma}{CFMA}{Compute-and-Forward Multiple Access}
\newacronym{d2d}{D2D}{Device-to-Device}
\newacronym{comac}{CoMAC}{Computation over Multiple Access Channels}

\newacronym{apc}{APC}{Average Power Constraint}
\newacronym{tpc}{TPC}{Total Power Constraint}
\newacronym{ipc}{IPC}{Individual Power Constraint}

\newacronym{mdf}{MDF}{Modified Detect-and-Forward}
\newacronym{maf}{MAF}{Modified Amplify-and-Forward}
\newacronym{dft}{DFT}{Discrete Fourier Transform}
\newacronym{pac}{PAC}{Parallel Access Channel}
\newacronym{dnf}{D\&F}{Decode-and-Forward}
\newacronym{cnf}{C\&F}{Compress-and-Forward}
\newacronym{anf}{A\&F}{Amplify-and-Forward}
\newacronym{cpnf}{CP\&F}{Compute-and-Forward}
\newacronym{stanc}{STANC}{Space-Time Analog Network Coding}
\newacronym{mtc}{MTC}{Machine Type Communication}

\newacronym{ia}{IA}{Interference Alignment}
\newacronym{ai}{AI}{Artificial Intelligence}
\newacronym{mwrc}{MWRC}{Multi-Way Relay Channel}
\newacronym{zf}{ZF}{Zero Forcing}
\newacronym{ldlc}{LDLC}{Low Density Lattice Codes}
\newacronym{svp}{SVP}{Shortest Vector Problem}
\newacronym{cpcnf}{CPC\&F}{Compute-Compress-and-Forward}
\newacronym{cran}{C-RAN}{Cloud-Radio Access Network}

\newacronym{bicmid}{BICM-ID}{Bit-Interleaved Coded Modulation with Iterative Decoding}
\newacronym{cmacr}{cMACr}{Compound Multiple Access Channel with a relay}

\begin{document}

	\title{The Magic of Superposition: A Survey on Simultaneous Transmission Based Wireless Systems}


	\author{Ufuk~Altun,~\IEEEmembership{Graduate Student Member,~IEEE},
        Gunes~Karabulut~Kurt,~\IEEEmembership{Senior Member,~IEEE}
        and Enver~Ozdemir,~\IEEEmembership{Senior Member,~IEEE}

}

\markboth{}%
{Altun \MakeLowercase{\textit{et al.}}: Bare Demo of IEEEtran.cls for IEEE Journals}

\maketitle

\begin{abstract}
	In conventional communication systems, any interference between two communicating points is regarded as unwanted noise since it distorts the received signals. On the other hand, allowing simultaneous transmission and intentionally accepting the superposition of signals and even benefiting from it have been considered for a range of wireless applications. As prominent examples, NOMA, joint source-channel coding, and the computation codes are designed to exploit this scenario. They also inspired many other fundamental works from network coding to consensus algorithms. Especially, federated learning is an emerging technology that can be applied to distributed machine learning networks by allowing simultaneous transmission. Although various simultaneous transmission applications exist independently in the literature, their main contributions are all based on the same principle; \textit{the superposition property}. In this survey, we aim to emphasize the connections between these studies and provide a guide for the readers on the wireless communication techniques that benefit from the superposition of signals. We classify the existing literature depending on their purpose and application area and present their contributions. The survey shows that simultaneous transmission can bring scalability, security, low-latency, low-complexity and energy efficiency for certain distributed wireless scenarios which are inevitable with the emerging internet of things (\acrshort{iot}) applications.

\end{abstract}


\IEEEpeerreviewmaketitle
\glssetwidest{BICM-IM}



\input{introduction}

\input{multipleaccess}

\input{multipleantenna}

\input{networkcoding}

\input{compalignment}

\input{functioncomputation}

\input{federatedlearning}

\input{spectrumsensing}

\input{detectionestimation}

\input{gossipconsensus}

\input{security}

\section{Conclusion}

The simultaneous transmission based communication techniques are the objective of this study. An extensive literature overview is presented on the wireless applications that exploit the interference of signals. These methods are grouped depending on their application areas and presented with detailed information. The wireless channel presents a natural weighted summation operation for the simultaneously transmitted inputs. This summation is exploited in various applications to perform desired tasks. From this perspective, simultaneous transmission techniques can be viewed as a channel manipulation that performs a specific function over the channel. The survey investigates these functions and their applications such as multiple access, network coding, detection, security etc. The studies in each application are listed along with their contributions and performance metrics.

\subsection{Challenges and Open Areas}

There are various challenges in simultaneous transmission networks that should be addressed and open to improvements. Most of the studies assume perfect \acrshort{csi} knowledge at both transmitters and receivers. Moreover, some of these studies are highly vulnerable to channel estimation errors which prevent implementation of these methods. For this reason, efficient and accurate \acrshort{csi} acquisition is essential and an open issue for simultaneous transmission methods. Various applications also require time synchronization among transmitting nodes.

The lack of perfect synchronization and \acrshort{csi} acquisition methods also effect existence of testbed implementation studies. Although there are \acrshort{afc} implementations via \acrshort{sdr} modules, the literature lacks testbed implementation studies on various applications such as network coding, multiple access or security.

Scalability and power constraints are other challenges for wireless networks. Although simultaneous transmission based methods improve scalability compared to the conventional pairwise methods, increasing network size still degrades the system performance in most applications. Moreover, the network size is also limited by the power constraints of the wireless channel.

Despite the challenges of the wireless environment, the simultaneous transmission techniques can bring scalability, security, bandwidth efficiency, low-complexity or low-latency for distributed networks. The future of communication is directed at the distributed networks, hence the simultaneous transmission can become a key component in future networks. The federated learning studies are already prominent examples that support this claim. Various application areas are still open to the improvements of simultaneous transmission. We believe that the hand-over process of mobile networks can be improved with simultaneous transmission. Moreover, implementation of simultaneous transmission methods for heterogeneous networks is another open area that is worth investigating.





\printglossary[style=alttree,type=\acronymtype,title=Abbreviations,nogroupskip, nonumberlist]


\bibliographystyle{IEEEtran}
\bibliography{ref}

%
%






\end{document}

%% file: introduction.tex
\section{Introduction}
\IEEEPARstart{T}{he} theory of communication, as presented in~\cite{Shannon1948} by Claude Shannon, considers the mathematical relation between a source and a destination. 
The theory identifies any signal between the source and the destination as noise. This assumption is still valid in advanced communication systems, especially in wireless communication networks that introduce destructive, as well as a limited medium to its users. Although its mobility feature attracts more and more users with each wireless network generation, limited resources present extreme design challenges since the wireless medium is shared by all of the participants. 

Some of the most critical design challenges are providing security, low latency and high communication rates and supporting a large number of users. From any of those aspects, the corresponding solution is sought, designed and tested for Shannon's pairwise communication model. Hence, the first step of the network design generally starts with virtually dividing the limited channel resources to each user and considering the crowded wireless channel as a combination of multiple subchannels between the pairs of nodes. For example, assigning orthogonal frequencies to different users is a practical and efficient approach to solve these challenges.

It is needless to say that all aspects we listed above are important for a wireless network, however their importance and the priority is closely related to the purpose of the network. With the emergence of \acrshort{iot} networks, we can add other design considerations to our list such as energy consumption or the system complexity. More importantly, an \acrshort{iot} network enables an immense range of wireless applications such that these aspects (and probably many more) are required in different and particular orders according to the application. For instance, security can be of high importance and the complexity is less for finance applications, while the mass sensor networks can require a low energy consumption and a high bandwidth efficiency.  

We believe that this diversity of wireless applications attracts the researchers' attention to the outside of the classical perspective. The physical layer based studies have already gained attention on this matter in order to help the wireless communication problems. One peculiar idea is related to the physical layer and the interference of signals. In the last two decades, the researchers have asked the question if the interference of signals can be beneficial. The question is closely investigated in several studies and positive answers were presented for a limited number of scenarios. These studies are mostly classified in the literature according to their application area, however their interactions with other studies regarding their unique channel model is usually overlooked. In this study, we consider the channel model aspect and survey the existing literature for the studies that exploit the superposition property of the wireless channel. The contributions of this survey can be listed as follows.

\begin{figure}[t]\center 
  \includegraphics[width=.96\linewidth]{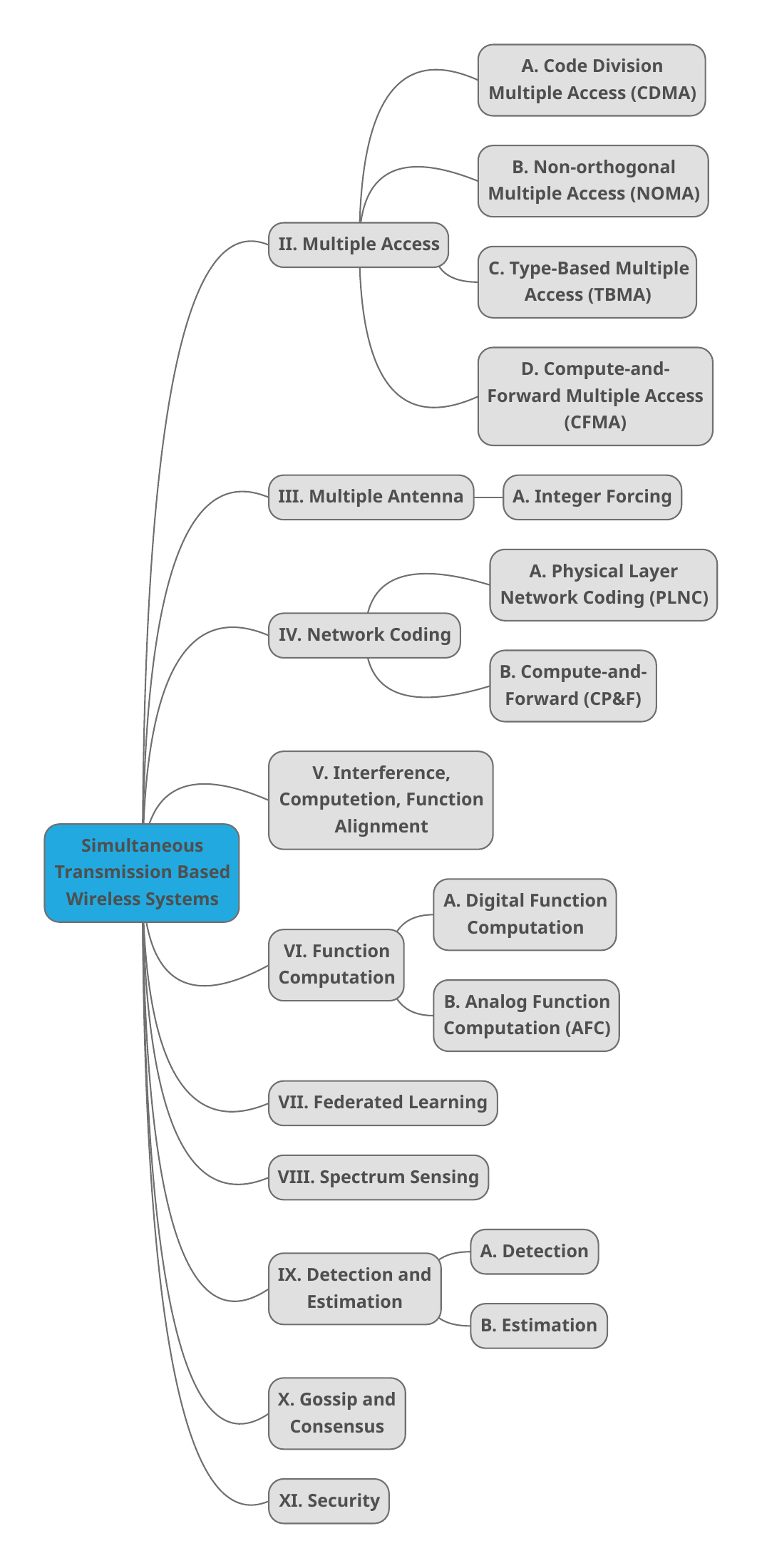}
  \caption{Structure of the survey as the main application areas of the simultaneous transmission based wireless communication techniques. } \label{fig:map} 
\end{figure}

\begin{itemize}
    \item \textit{The studies in the literature that exploit the superposition property of the wireless channel are reviewed:} A variety of wireless communication techniques exist in the literature (Fig.~\ref{fig:map}) with this apparent and unique connection (superposition property). Yet the literature lacks a comprehensive guide that highlights this connection to the readers. This survey presents a guide for both beginner and experienced readers that are interested in the benefits of simultaneous transmission.  
    \item \textit{The studies are presented along with their system model, contributions and performance metrics:} The simultaneous transmission methods are grouped depending on their application areas. The methods are introduced with brief explanations and studies are investigated for their contributions and performance metrics.
\end{itemize}

\subsection{Boundaries of Our Scope}

Since we consider red{interfering} signals, we are naturally interested in the multiple users and the techniques for accessing the wireless channel. It should be pointed out that the answers to all our questions are not included within Shannon's communication model regarding the interference policy. Here our boundaries may be perplexing since there are certain techniques in the literature that accept the interference of other signals, however still interested in the pairwise communication without benefiting from interference. 

The general resources of a wireless network is use of power in the time and the frequency plane that is suitable for distribution among the users. However, communication technologies make use of other resources such as code or energy to divide among the users. The users in these studies occupy the same time and frequency blocks, yet they are distributed with another resource. The code division multiple access (\acrshort{cdma}) is an example that enables controlled interference and uses the code dimension for multiple access. The spatial diversity of the multiple input multiple output (\acrshort{mimo}) networks can be given as another example that does not prohibit interference. However, \acrshort{mimo} technology should be considered under pairwise communication since its interference comes from the signals that are known and managed by a single user.

These examples raise the question of whether our scope is limited to any technique that enables the interference of signals (i.e. uses the same time and frequency slots) or limited to a more narrow scenario. We could not answer this question with a single border since one of them would draw an incomplete map on the main idea and the other one would be too extensive to cover and also include unrelated parts. For this reason, our scope also includes the major techniques (e.g. \acrshort{cdma}, \acrshort{mimo}) that enable the interference of signals but not directly got benefit from it. However, we briefly illustrate their relation to the interference and we refer the readers to more detailed resources on that matter. 

Our main focus is on the techniques that intentionally exploit the interference of signals and get a direct benefit from it. Multiple users in the techniques that we present here use the same time and frequency blocks for communication and do not aim to distinguish between the individual information. Instead, these methods are only interested in the superimposed form of the signals. There are various techniques in the literature and before giving a general map of these techniques, first we would like to clarify some of the concepts that we use to avoid any ambiguity.

We define the channel model as the wireless multiple access channel (\acrshort{wmac}). The multiple access term states the fact that the simultaneously transmitted electromagnetic waves combine with each other. In general, this statement only covers the time dimension, i.e. the waves from other frequencies also merge over the \acrshort{wmac}. However, it is straightforward to separate the superimposed waves in the frequency dimension at the receiver. For this reason, we extend our definition of simultaneous transmission and include the frequency dimension. Hereafter, we refer to simultaneous (concurrent) transmission to cover both time and frequency domains, i.e. signals are transmitted at the same time and frequency block. Similarly, we refer to the simultaneously transmitted signals as the \textit{superpositioned} or \textit{superimposed} signals. Furthermore, we use \textit{superposition} or \textit{simultaneous transmission} to indicate the methods that we are interested in.

\subsection{A Map of the Simultaneous Transmission Based Communication Techniques}

There are several application areas of \textit{simultaneous transmission}-based techniques for different purposes. Each application has its own design parameters and performance metrics, for instance distributed detection problems usually require a high detection probability while security applications are interested in secrecy rates. As a result, this survey assumes a unique point of view that connects various problems and application areas of the wireless communication networks. However, the \textit{superposition} property of the \acrshort{wmac} implies several common grounds as follows:
\begin{itemize}
    \item Its distributed character: All applications include decentralized multiple nodes. As a result, scalability, energy efficiency and complexity are often a design concern.
    \item Its wireless character: The users communicate over the wireless channel. The amount of channel state information (\acrshort{csi}) knowledge at the users and the channel characteristics (e.g. fading) are important concerns. 
\end{itemize}

The structure of this survey (the application areas of the simultaneous transmission based techniques) is illustrated in Fig.~\ref{fig:map}. The multiple access and the multiple antenna applications are the early examples that the \textit{superposition} of the simultaneously transmitted signals is observed at the receiver. \acrshort{cdma} and non-orthogonal multiple access (\acrshort{noma}) techniques are essentially based on pairwise communications. For this reason, transmitters use a third resource (other than time and frequency) to distinguish between the individual messages, e.g. the \acrshort{cdma} or \acrshort{noma} users uniquely encode their messages as a function of a code block or a power level respectively. On the other hand, the traditional \acrshort{mimo} applications benefit from the superimposed signals to gain diversity although the network is not distributed. 

\begin{figure}[t]\center 
  \includegraphics[width=1\linewidth]{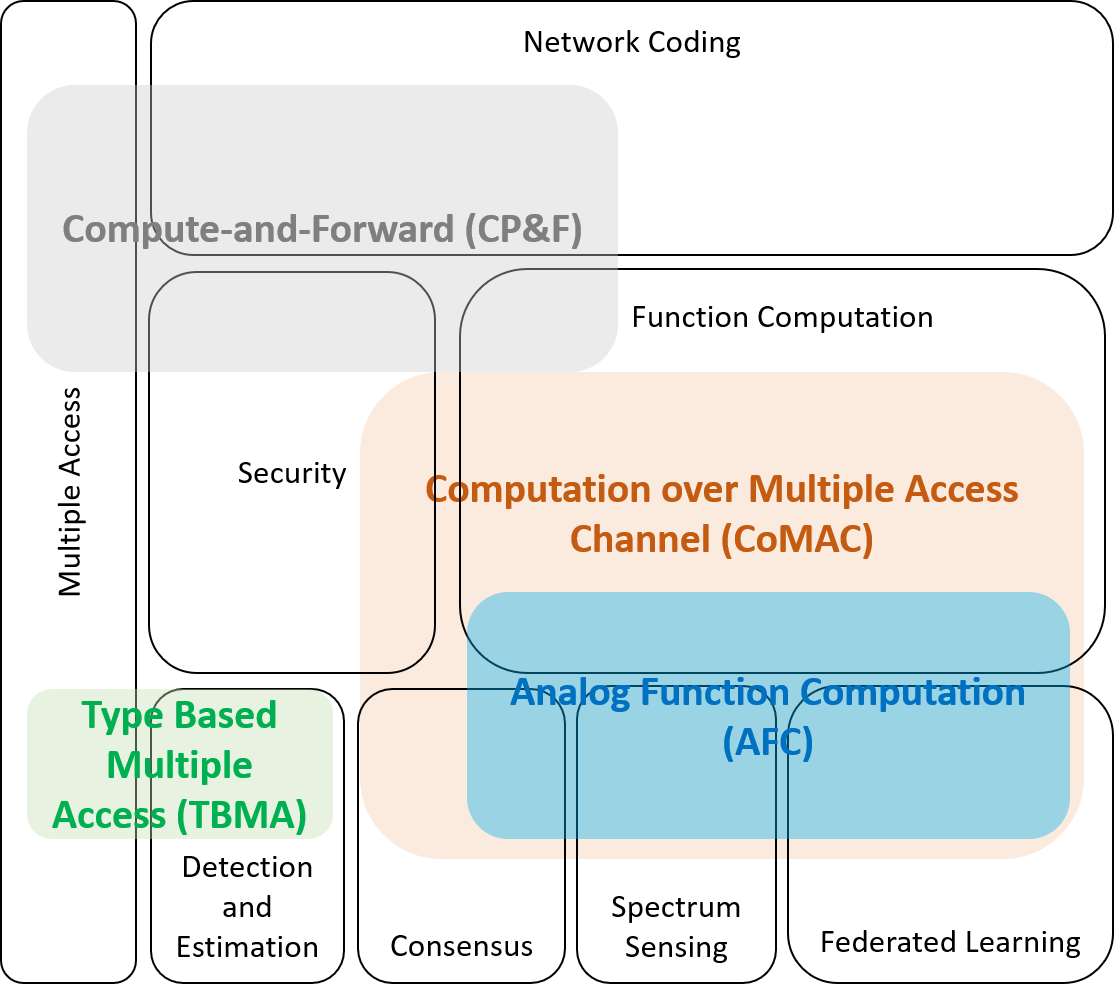}
  \caption{An illustration of the featured methods and their applications areas. Colourless rectangles represent application areas where coloured titles are the featured methods. A featured method is well used in the application areas where their rectangles intersect.} \label{fig:featured}
\end{figure} 

\begin{figure*}[t]\center 
  \includegraphics[width=1\linewidth]{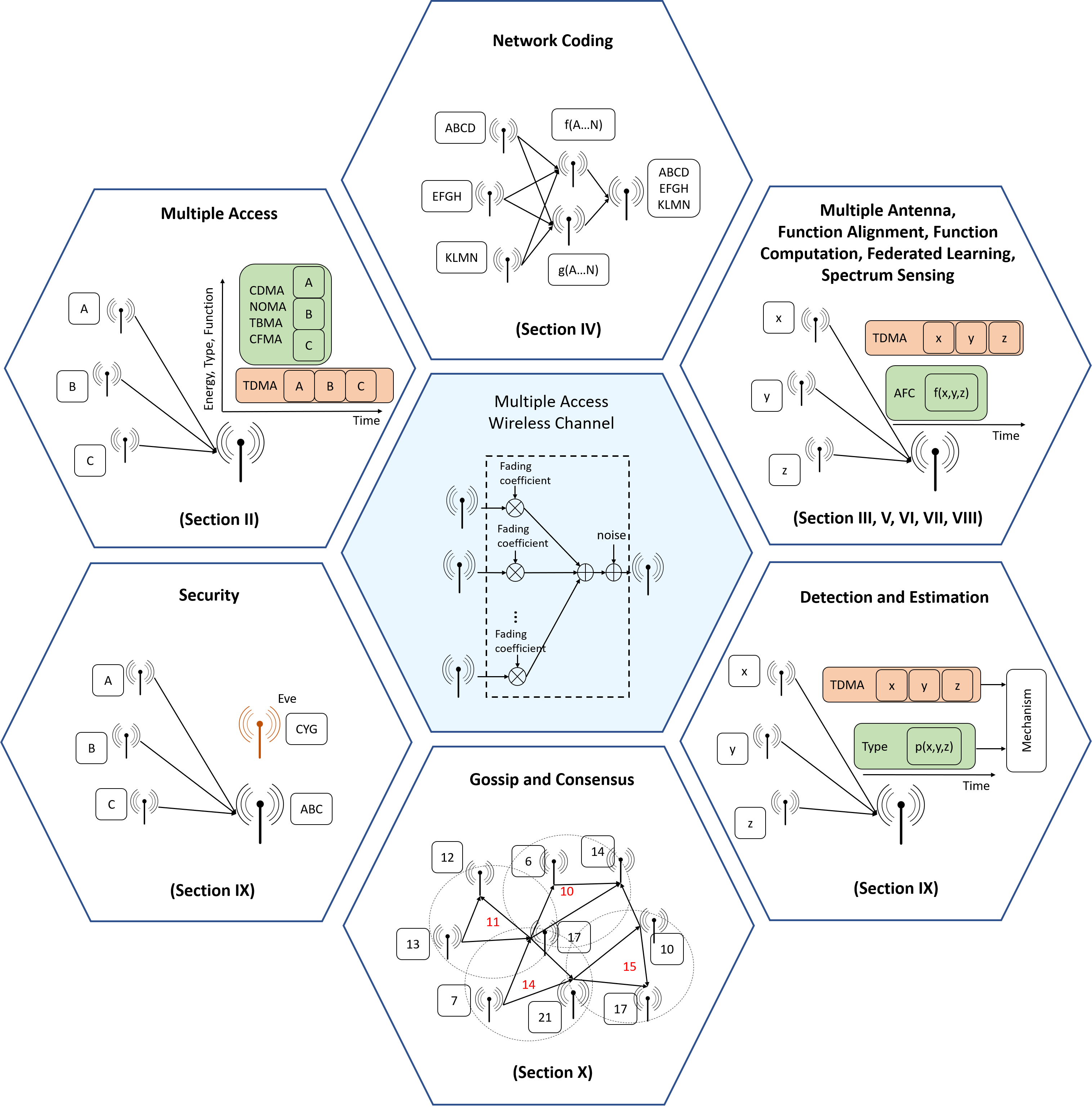}
  \caption{An illustration of the relationship between the applications and their network model.} \label{fig:relation}
\end{figure*} 

Joint source-channel coding proposed in~\cite{Gastpar2003} has created a paradigm shift on the multi-user communication models by considering the joint optimization of communication and computation aspects. Over the years, the paradigm has been shaped into several core models such as type-based multiple access (\acrshort{tbma})~\cite{Mergen2006}, computation over multiple access channel (\acrshort{comac})~\cite{Nazer2007}, compute-and-forward (\acrshort{cpnf})~\cite{Nazer2011} and analog function computation (\acrshort{afc})~\cite{Goldenbaum2013}. These studies influenced novel techniques in the past decade and appeared at various application areas from network coding to federated learning. An illustration of these techniques and their general application areas are given in Fig.~\ref{fig:featured}.

\acrshort{cpnf} is mainly investigated for relaying and network coding purposes to improve the network capacity (i.e. computation rate), however it has also been extended to multiple access applications (\acrshort{cfma}~\cite{Sula2019}) to reduce complexity and security applications (as in~\cite{Vatedka2015}) to improve secrecy rate. The computation alignment and the function alignment studies which are influenced by \acrshort{cpnf} and \acrshort{comac} are proposed to reach the computing capacity in multiple antenna networks.

\acrshort{comac} inspired various digital and analog function computation methods that aims to allow low-latency, low-bandwidth computations in distributed wireless networks. The main idea behind the \acrshort{comac} is later implemented in spectrum sensing, federated learning and consensus algorithms to improve network efficiency. Especially, federated learning algorithms are one of the latest and most promising application area of the simultaneous transmission. With the help of \acrshort{comac}, federated learning algorithms can decentralize machine learning systems and improve energy or time efficiency of the system. \acrshort{tbma} is primarily used for distributed detection and estimation applications to design scalable and energy-efficient models.

From an intriguing perspective, all these models and applications can be viewed as a particular function that manipulates the wireless channel to perform a given task. For example, multiple access algorithms aim to transfer the transmitted data to a destination. Eventually, simultaneous transmission manipulates the channel to perform a function which both inputs and the outputs are the transferred data. However in the detection algorithms, outputs are not data, instead they are the test statistics. Hence, it can be said that a detection algorithm that uses simultaneous transmission manipulates the wireless channel to perform a function that inputs are the sample set and the output is the test statistics. The corresponding input-output relationships are illustrated in Fig.~\ref{fig:relation}. 

The security applications consider transferring data to the receiver without leaking information to the eavesdroppers. This approach manipulates the \acrshort{wmac} to transfer the initial data to the receiver such that the received signal is only meaningful at the destination. Network coding applications consider transferring data in layered network structures. The simultaneous transmission models propose a cascade of functions (composite functions) (relays) that both the initial inputs and the final outputs are the data. Function computation, federated learning, spectrum sensing algorithms match the \acrshort{wmac} with particular functions that is unique to the purpose of the application. Moreover, gossip and consensus algorithms also consider the topology of the network by using subgroups or subfunctions in the network, i.e. composite functions.

In this survey, we consider wireless communication techniques that benefit from simultaneous transmission and we classify the existing literature according to their application purposes. We draw a map of the existing literature and provide introductory information on each application (Fig.~\ref{fig:featured}). We also give details of these studies and investigate their contributions as well as their performance metrics. 

The paper is organized as follows. In Section II, multiple access methods that benefit from the simultaneous transmission are presented. Section III is dedicated to the multiple antenna algorithms. Network coding studies are investigated in Section IV. Interference alignment, computation alignment and function alignment studies are examined in Section V. Digital and analog function computation applications are presented in Section VI, and the federated learning studies are presented in Section VII. In Section VIII, spectrum sensing models are presented. Section IX is devoted to the detection and estimation studies, and Section X includes the gossip and consensus studies. Lastly, the security applications are given in Section XI. The paper is concluded in Section XII.

%% file: multipleaccess.tex
\section{Multiple Access} \label{sec: multipleaccess}

Wireless channel has limited resources (signaling dimensions) such as frequency, time, or space and accessing to the channel requires a portion from each resource. Multiple access methods aim to allocate these resources to multiple users efficiently~\cite{goldsmith2005}. Our interest, simultaneous transmission, enables the dedication of all frequency and time domains to all users. As a result, simultaneous transmission for the purpose of multiple access falls under the category of our interest. However, known multiple access methods such as \acrshort{cdma} and \acrshort{noma} divide another dimension to its users rather than exploiting the superposition of the signals. For this reason, we find it more adequate to give elementary information on these methods and refer the readers to proper references. In this section, multiple access methods that enable users to simultaneously transmit their messages are presented.

\subsection{Code Division Multiple Access (\acrshort{cdma})}
\begin{figure}[t]
\begin{subfigure}[t]{.24\textwidth}
  \centering
  \includegraphics[width=.9\linewidth]{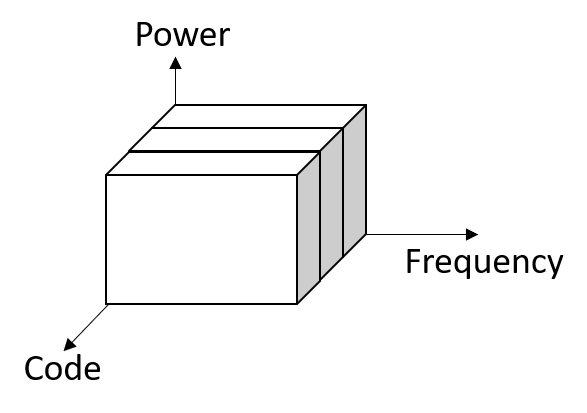}  
  \caption{}
  \label{fig:cdma} 
\end{subfigure}
\begin{subfigure}[t]{.24\textwidth}
  \centering 
  \includegraphics[width=.9\linewidth]{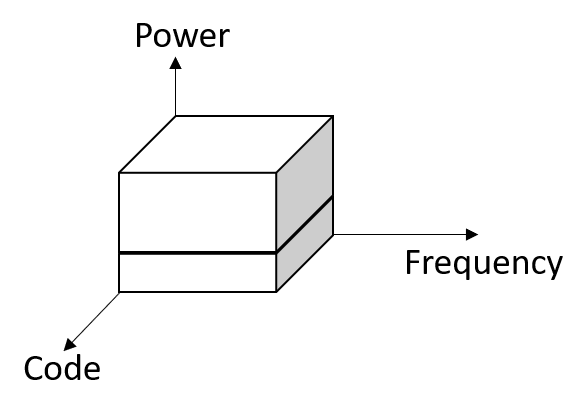}  
  \caption{}
  \label{fig:noma}
\end{subfigure}
\caption{Distribution of resources in multiple access methods of our interest. (a) CDMA (b) NOMA}
\label{fig:multpleaccess}
\end{figure}

In \acrshort{cdma}, each user is assigned with a spreading code to distinguish users from each other. Therefore, the channel can be used by all users in the same time period and bandwidth. Especially in the uplink scenario, in which multiple users transmit simultaneously, base station receives a combination of signals from all users. If the codes are orthogonal, despreading the received signal with the corresponding code outputs the information signal of the corresponding user. Fig.~\ref{fig:cdma} illustrates the resource distribution of CDMA technique. As seen from the figure, each user occupies a large and equal bandwidth and usually, a power balance is required. Non-orthogonal codes are also used in CDMA to flex the synchronization problem and support more users, however removing orthogonality adds interference to the system.    

In~\cite{PRASAD1998a}, extensive information is given about the working principles of the CDMA as well as an introduction to the various versions of the CDMA. An extended CDMA method, complementary code based MIMO CDMA, which aims to revive CDMA in the next-generation systems are investigated in~\cite{Sun2015}.

\subsection{Non-Orthogonal Multiple Access (\acrshort{noma})}
\acrshort{noma} is an emerging multiple access method that is based on the distribution of power domain\footnote{NOMA is usually referred for power domain distribution. However, the name also suggests a category that includes any non-orthogonal method (i.e. two categories; orthogonal and non-orthogonal multiple access methods). We only consider the power domain NOMA.}. NOMA can be viewed as a complementary method for the existing multiple access techniques since the power domain is not suitable to support more than two or three users. As a result, the distribution of another resource is required. However, \acrshort{noma} based systems still remain in our scope since \acrshort{noma} enables the transmission of two users simultaneously. Resource distribution of \acrshort{noma} is given in Fig.~\ref{fig:noma}. The power is intentionally distributed to users with a level difference. The model depends on the successive interference cancellation (\acrshort{sic}) such that the receiver is able to detect the higher level or lower level signal and then cancel it. 

\acrshort{noma} is an emerging multiple access method and attracts attention for the next generation communication systems. We refer the readers to~\cite{Islam2017} and~\cite{Dai2018} for comprehensive information on \acrshort{noma}. In~\cite{Makki2020} an overview on the current challenges of \acrshort{noma} (also the rejection of \acrshort{noma} in 5G standards) and possible solutions are analyzed. Additionally, performance of \acrshort{noma} is investigated with computer simulations.

\subsection{Type-Based Multiple Access (\acrshort{tbma})} 
\acrshort{tbma} is a unique method by means of channel-user relationship since the users in \acrshort{tbma} does not aim to communicate with the destination individually. Instead, a data statistics (type) of all the users is transferred to the destination~\cite{Jeon2011}. It should be noted that the individual data is not reconstructed at the receiver and \acrshort{tbma} is not suitable for classical wireless communications. However, it reduces the network latency and shows huge potential in sensor networks that only the total data statistics is desired. Specifically, \acrshort{tbma} is widely used for detection and estimation purposes.

\acrshort{tbma} is not widespread in the literature and requires a broader explanation of its mechanism. For this reason, we dedicate this section solely to the explanation of the \acrshort{tbma}. The literature review of \acrshort{tbma} is given in the detection and estimation section. 

Consider a network that consists of $n$ users that aims to transfer messages (e.g. sensor readings) $X_i, X_2,...,X_n$ to a fusion center (\acrshort{fc}). The $i^{th}$ user in the network assigns its message $X_i$ to a waveform $s_{x_i}$ that is chosen from an orthonormal waveform set $\{ s_1, s_2, ... s_k \}$. When each user in the network simultaneously transmit its data with energy $E$, the \acrshort{fc} obtains the following expression~\cite{Mergen2007}
\begin{equation}
    z = \sum_{i=1}^{n} h_i \sqrt{E} s_{x_i} +w,
\end{equation}
where $h_i$ and $w$ are the channel and fading coefficients respectively. Assuming the channel gains are inverted, the signal at the \acrshort{fc} becomes
\begin{equation}
    z = \sum_{j=1}^{k} \sqrt{E} N_j s_j +w,
\end{equation}
where $N_j$ is the number of sensor readings that is mapped into waveform $s_j$. Note that each member of the message set is mapped into a waveform and the transmitted signals are superimposed over the channel. As a result, for each waveform, the \acrshort{fc} receives the number of users that transmitted the corresponding waveform (i.e. the FC obtains histogram (type) of the sensor readings).

\subsection{\acrfull{cfma}}

\acrshort{cpnf} is a relaying and network coding method that is based on the simultaneous transmission of signals. The objective of the \acrshort{cpnf} is to efficiently transfer the messages of multiple sources to a receiver with the help of multiple relays. In essence, \acrshort{cpnf} transfers a function of the source messages to each relay by exploiting the superposition property. The relays are unable to decode the individual messages since each of them obtains a single function that contains multiple messages (unknown parameters). The relays forward their functions to the receiver and the receiver can reconstruct the individual messages by solving the functions for the unknown parameters. Contrary to other network coding algorithms, the communication phase between the sources and the relays takes place at the same time slot and the bandwidth in \acrshort{cpnf}, hence it offers efficiency on the spectrum and latency. 

Compute-and-forward multiple access (\acrshort{cfma}) is inspired by the \acrshort{cpnf} and instead of relaying the messages, it enables direct access to the channel by exploiting the superposition property. The \acrshort{cfma} is proposed in~\cite{Zhu2017a} by Zhu and Gastpar for the networks that two users aim to access to a single receiver. The main idea behind \acrshort{cfma} is based on the same coding and decoding structure as the \acrshort{cpnf} which yields a function of the messages at the receiver. However, two users directly communicate towards a receiver without multiple relays. Two functions are required at the receiver to solve the messages. For this reason, \acrshort{cfma} also uses successive cancellation decoding to obtain the coefficients of the second function. The authors also examine the more than two users scenario in~\cite{Zhu2017a} and low density parity check (\acrshort{ldpc}) coded \acrshort{cfma} is considered in~\cite{Sula2017} and~\cite{Sula2019}. 

In conclusion, CDMA and NOMA methods accept the interference of multiple users in their nature. However, the methods still aim to reconstruct each signal at the receiver by using another distinguishing factor such as signal power or code. In TBMA and CFMA, signal superposition is purely accepted in order to improve bandwidth and time efficiency. Major drawbacks such as imperfect channel knowledge or synchronization prevents TBMA and CFMA from successful implementations. Although signal superposition can bring mass scalability to a multiple access schemes, it requires error free hardware to be applicable. In the following section, multiple antenna techniques that benefit from the simultaneous transmission are presented.

%% file: multipleantenna.tex
\section{Multiple Antenna}

Multiple antenna techniques became beneficial in the communication networks as a result of the developments in both the antenna technology and the processing capacity. \acrshort{mimo} structure is proven to improve the multiplexing capability of the single antenna networks as well as increasing their diversity gain~\cite{Mietzner2009}. \acrshort{mimo} presents a unique case for the scope of this survey since the communication is between pairwise nodes, e.g. the antennas are controlled by the same source. The fundamental advantage of \acrshort{mimo} is the diversity gain which results from the superposition of the signals that are transmitted from multiple antennas. As a result, \acrshort{mimo} networks draw our attention in the sense that the superposition property is exploited to reduce the error rates or improve the bit rates. On the other hand, the conventional \acrshort{mimo} is an extensive topic and the superposition property is just a tool in \acrshort{mimo} studies which leads to numerous results~\cite{Agiwal2016}. Also, it is already well presented in the literature and we believe that the relation between \acrshort{mimo} and the superposition property can be better observed from the existing studies such as~\cite{Castaneda2017, Xu2017}. Further reading on \acrshort{mimo} can be found in~\cite{Chen2017} for its security applications and in~\cite{Yang2015} for its challenges and future. In addition to its conventional perspective, there also exists studies that exploit the superposition property in a unique way with multiple antennas. The integer-forcing receivers~\cite{Zhan2010, Zhan2012} are an example of these studies and we would like to mention its architecture and its difference from the traditional studies.

\subsection{Integer-Forcing Architecture}

An effective channel matrix can be defined and used for the analysis of linear \acrshort{mimo} receivers. In the traditional sense, the effective channel matrix should be matched with an identity matrix in order to recover the messages of each antenna. However, integer-forcing receivers match the effective channel matrix with integer value matrices as proposed in~\cite{Zhan2010}. Inspired by \acrshort{cpnf}~\cite{Nazer2011}, nested lattice codes are used for communication in order to obtain functions with integer coefficients. After matching with integer values, the receiver can solve the effective channel matrix and obtain messages if the matrix is full rank.

The integer-forcing receiver is extended to mitigate the external signal interference in~\cite{Zhan2011}. In addition to the results given in~\cite{Zhan2010} that integer-forcing receivers obtain the inputs with integer coefficients, the authors later observe in~\cite{Zhan2011} that the integer values can be also controlled to mitigate the external interference by considering the interference space. The results show that the proposed receiver presents significant gain over the traditional linear \acrshort{mimo} receivers.

Successive interference cancellation (\acrshort{sic}) technique is adapted to the integer-forcing receivers in~\cite{Ordentlich2013} as successive integer-forcing. The results indicate that the successive integer-forcing receiver can achieve the channel's sum capacity and outperform the traditional linear receivers with \acrshort{sic} in certain scenarios. 

Multiple antenna schemes proves that superposition of signals can be enabled to improve diversity gain of a communication system. Especially the integer-forcing architecture exploits lattice codes to solve transmitted signals with further gain. However, solving the effective channel matrix is not always easy or possible. Also, imperfect channel estimation is the major drawback of integer-forcing architecture.

%% file: networkcoding.tex

\section{Network Coding} \label{sec: networkcoding}

\begin{figure}[t]
\begin{subfigure}[t]{.24\textwidth}
  \centering
  \includegraphics[width=.8\linewidth]{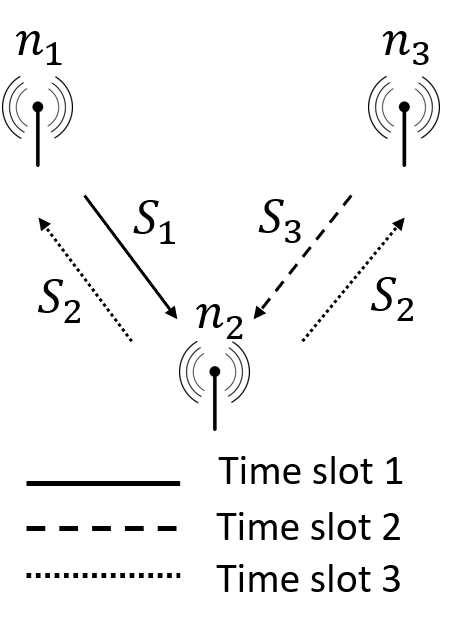}
  \caption{}
  \label{fig:nc}
\end{subfigure}
\begin{subfigure}[t]{.24\textwidth}
  \centering
  \includegraphics[width=.8\linewidth]{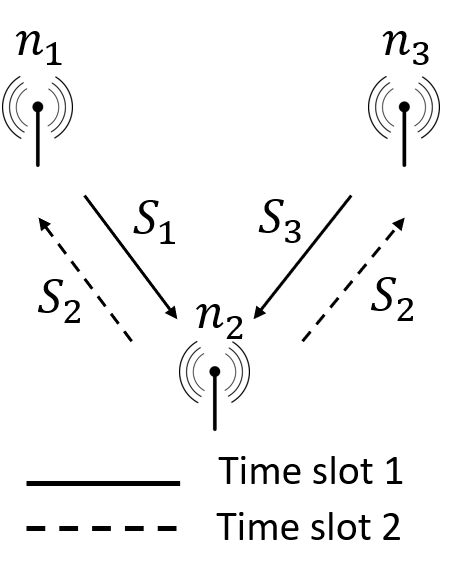}  
  \caption{}
  \label{fig:plncc}
\end{subfigure}
\caption{Information flow comparison between traditional network coding and \acrshort{plnc}~\cite{Zhang2006}. (a) Traditional network coding (b) \acrshort{plnc}}
\label{fig:plnc}
\end{figure}

Source and channel coding are essentially concerned with the communication between two nodes. Specifically, they improve the capacity and error performance of pairwise communication respectively. On the other hand, network coding benefits from the architecture of the network in order to improve its capacity, efficiency and security~\cite{Bassoli2013}. Network coding is interested in the information flow between nodes. The fundamental idea behind the network coding can be seen in Fig.~\ref{fig:nc}. The nodes $n_1$ and $n_3$ aims to exchange information through an in-between node $n_2$ in the given network. Without network coding, the messages $S_1$ and $S_3$ of the nodes $n_1$ and $n_3$ would require a total number of four pairwise hops, hence four-time slots. A simple network coding algorithm can be applied, as given in Fig.~\ref{fig:nc}, to reduce the required time slots. After obtaining $S_1$ and $S_3$ sequentially, the relay node $n_2$ computes $S_2 = S_1 \oplus S_3$ and broadcasts it. Since the nodes know their initial messages, $n_1$ and $n_3$ can extract the unknown message from $S_2$. As a result, the network gains a time slot with the broadcast of the relay node.

Network coding is one of the most effective and intriguing applications of the simultaneous transmission since the multiple access nature of the channel presents unique opportunities for the code design. These studies are specifically called physical layer network coding (\acrshort{plnc}) which outperforms the traditional network coding in certain scenarios~\cite{Hu2010}. We consider the \acrshort{plnc} in two subsections since one study, the \acrshort{cpnf}~\cite{Nazer2011}, made a name for itself and requires special attention.

\begin{figure*}[t]
\begin{subfigure}[t]{.32\textwidth}
  \centering
  \includegraphics[width=1\linewidth]{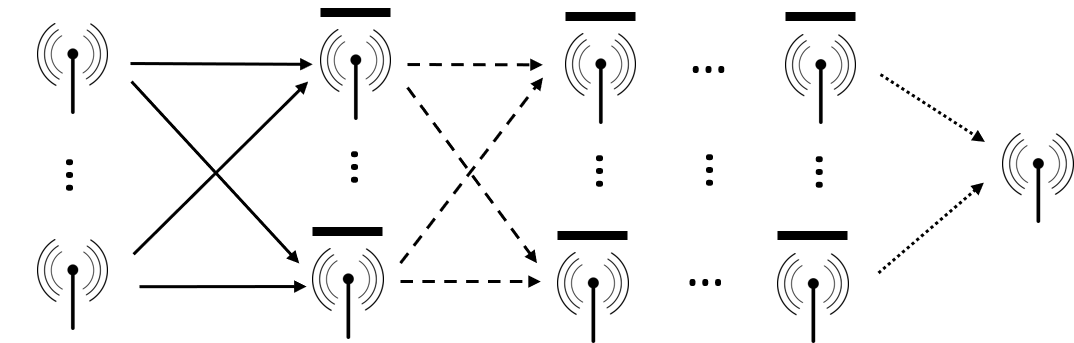} 
  \caption{}
  \label{fig:multihop}
\end{subfigure}
\begin{subfigure}[t]{.24\textwidth}
  \centering
  \includegraphics[width=.8\linewidth]{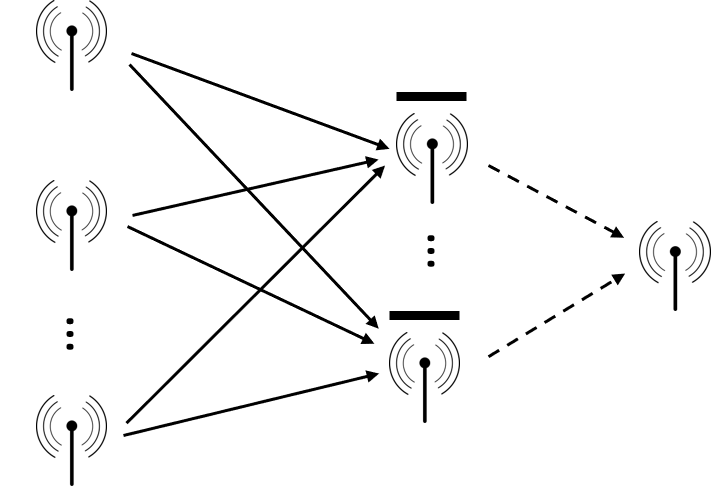}  
  \caption{}
  \label{fig:multinode} 
\end{subfigure}
\begin{subfigure}[t]{.18\textwidth}
  \centering
  \includegraphics[width=.8\linewidth]{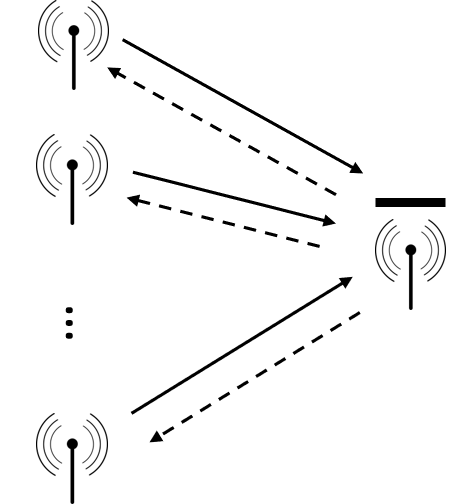}  
  \caption{}
  \label{fig:multiway} 
\end{subfigure}
\begin{subfigure}[t]{.2\textwidth}
  \centering
  \includegraphics[width=1\linewidth]{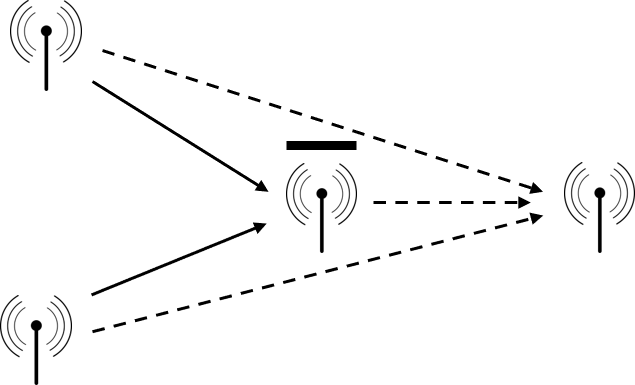}  
  \caption{}
  \label{fig:cooperative} 
\end{subfigure}
\caption{Several network topologies that are commonly used in simultaneous transmission based techniques. Lines over the nodes represent relays. (a) Multi-hop (b) Multi ($K$) transmitter multi ($L$) relay (c) Multi-way relay (d) Cooperative}
\label{fig:relaystructures}
\end{figure*}

The network architecture is an important parameter in the investigation of these studies and some commonly used architectures are illustrated in Fig.~\ref{fig:relaystructures}. Also, the studies are classified according to their network architectures in Table~\ref{tab:plnc} along with their performance metrics. In the coming part, we briefly explain the main ideas behind these methods and investigate the studies that exist in the literature.

\subsection{\acrfull{plnc}}

The \acrshort{plnc} is proposed in 2006 by Zhang \textit{et al.}~\cite{Zhang2006}. The proposed network is based on the superposition of the signals to reduce the required time slots. A simple three-node example of the \acrshort{plnc} is given in Fig.~\ref{fig:plncc}. Without the conventional network coding, it is obvious that the uncoded scheme requires a time slot for each transmission and the conventional network coding reduces the required number of transmission by enabling broadcast at the relay. 


\acrshort{plnc} further improves this situation by accepting the superposition of the messages from $n_1$ and $n_3$ to the relay node as seen in~\ref{fig:plncc}. In \acrshort{plnc} scenario, relay node receives the superimposed message and can not extract $S_1$ and $S_3$. However \acrshort{plnc} allows $n_1$ and $n_3$ to decode the superimposed message after $n_2$ broadcasts it. Later, this basic example is extended to larger networks and different coding schemes as in~\cite{Katabi2007} that a packet-based \acrshort{plnc} architecture is proposed and implemented. The study uses a testbed of computers and establishes a proof of concept for the \acrshort{plnc}.

Analog version of the \acrshort{plnc}~\cite{Zhang2006} is proposed in~\cite{Katti2007}. The authors simply consider the superposition of the signals instead of packets. Also, the relay uses amplify-and-forward (\acrshort{anf}) to transfer the superimposed signal. After presenting their results, the study also verifies them by implementing the proposed method with software defined radio (\acrshort{sdr}) modules. 

Another analog \acrshort{plnc} scheme, space-time network coding (\acrshort{stanc}), is proposed in~\cite{Amah2011}. The study considers a non-regenerative multi-way relay network that multiple nodes (equipped with single antenna) exchange information through a single relay (equipped with multiple antennas) in stationary and non-stationary channels. The \acrshort{stanc} is proposed for the stationary case and an alternative solution, the repetition transmission, is suggested for the non-stationary channels. Achievable sum rates of these two models are calculated and verified with simulations. The simulations also included a comparison with the zero-forcing (\acrshort{zf}) and maximization of \acrshort{snr} beamforming models and it is shown that the \acrshort{stanc} outperforms other models regarding the sum rates. The network model in~\cite{Wei2013} is customized to the networks where multiple nodes transfer information to a single receiver both directly and with the help of a relay node (equipped with multiple antennae). The sum-rate performances and error rate performances are evaluated with simulations for the cases of direct transmission, analog network coded transmission and the \acrshort{stanc} transmission.

In~\cite{Gunduz2010}, a network that consists of a single relay, two transmitters and two receivers (compound multiple access channel with a relay (\acrshort{cmacr})) is considered. The relay node is assumed to have cognitive capabilities and able to include its own message to the received message before forwarding. The study investigates the achievable rate regions of three relay methods; decode-and-forward (\acrshort{dnf}), compress-and-forward (\acrshort{cnf}) and lattice coded (\acrshort{cpnf}) schemes. Also, a special case is examined where the \acrshort{cmacr} network does not allow cross-reception (i.e. one of the sources always connects to the receiver via the relay, not directly). In this scenario, the modulo sum of the source messages is computed with the lattice codes and compared with the \acrshort{dnf} and \acrshort{cnf} schemes.

A multi-hop network includes multiple layers of relays between the sources and the destiny as illustrated in Fig~\ref{fig:multihop}. In~\cite{Xu2012}, a cross-layer strategy is followed for a multi-hop \acrshort{plnc} network to design the efficient routing paths. In~\cite{Burr2014}, algebraic frameworks are considered for the design of multi-hop \acrshort{plnc} networks. 

The compatibility and the performance of the \acrshort{plnc} with error correction codes are investigated in~\cite{Al-Rubaie2013}. Turbo codes, \acrshort{ldpc} codes and bit-interleaved coded modulation with iterative decoding (\acrshort{bicmid}) are simulated in a \acrshort{plnc} based network. The results showed that the \acrshort{plnc} reduces the bit error rate (\acrshort{ber}) performance of all three channel coding schemes.

In~\cite{Hayashi2019}, a secure \acrshort{plnc} scheme is designed. The method is inspired by the forwarding algorithm given in~\cite{Ren2017} (based on \acrshort{cpnf}). The study analyzes two networks with a butterfly topology and a three source topology. Also, it is shown that the secure \acrshort{plnc} outperforms the secure network coding such as given in~\cite{Cai2011}.

\begin{table*}[t]  \caption{An overview of \acrshort{plnc} studies.}
  \centering 
 \begin{tabular}{ p{2.5cm}|p{0.7cm}|p{1.7cm}|p{7cm}|p{3cm}  }
 Author  &  Year  & Network model & Contribution & Performance metric  \\

  \hline \hline
    Zhang \textit{et al.}~\cite{Zhang2006} & 2006  & Two-way (multi-hop) & Traditional \acrshort{plnc} is introduced. & \acrshort{ber} \\ \hline
    
    Katti \textit{et al.}~\cite{Katabi2007} & 2008 & Multi-way & Throughtput improved with \acrshort{plnc}. & Throughtput gain \\ \hline

    Katti \textit{et al.}~\cite{Katti2007} & 2007 & Two-way & Analog \acrshort{plnc} is introduced. & Network throughput
    \\ \hline

    Amah and Klein~\cite{Amah2011} & 2011 & Multi-way & \acrfull{stanc} introduced. & Sum rate \\ \hline

    Wie and Chen~\cite{Wei2013} & 2013 & Cooperative & \acrshort{stanc} adapted to multi-way cooperative networks. & Sum rate \\ \hline

    Gündüz \textit{et al.}~\cite{Gunduz2010} & 2010 & Cooperative (\acrshort{cmacr}) & Lattice codes used for \acrshort{plnc} in cooperative networks. & Achievable rate region \\ \hline

    Xu \textit{et al.}~\cite{Xu2012} & 2012 & Multi-hop & \acrshort{plnc} implemented to multi-hop networks with a cross-layer design. & Network throughput \\ \hline

    Burr and Fang~\cite{Burr2014} & 2014 & Multi-hop & Algebraic constructs are used in the network design. & Network throughput \\ \hline
    
    Al-Rubaie \textit{et al.}~\cite{Al-Rubaie2013} & 2013 & Two-way & \acrshort{ldpc} and Turbo codes are compared. & BER \\ \hline

    Hayashi~\cite{Hayashi2019} & 2019 & Cooperative (Butterfly) & \acrshort{plnc} and NC are compared. & Number of time spans \\

  \end{tabular}

  \label{tab:plnc}
\end{table*}

\begin{figure}[t]\center
  \includegraphics[width=1\linewidth]{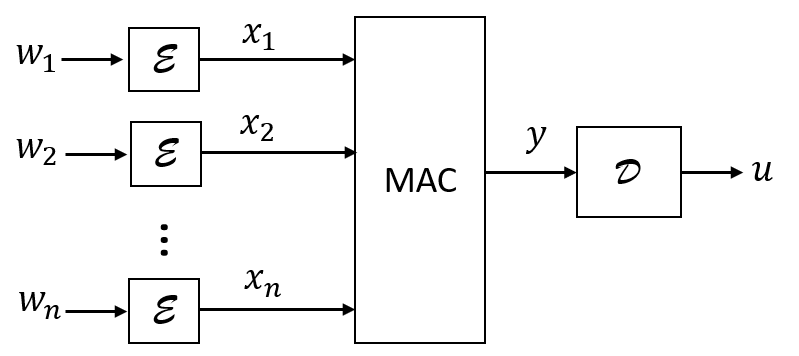}
  \caption{The network model of the \acrshort{cpnf} method. $\mathcal{E}$ and $\mathcal{D}$ symbolize encoder and decoder, respectively.}
  \label{fig:cpnf}
\end{figure}

\subsection{Compute-and-Forward (\acrshort{cpnf})}

\acrshort{cpnf} is proposed by Nazer and Gastpar in~\cite{Nazer2011} as a relaying method and draw the attention of numerous researchers throughout the years. \acrshort{cpnf} is inspired from the lattice codes (structured codes) that is previously used in~\cite{Nazer2007, Nazer2007f} and~\cite{Nazer2007g}. A lattice $\Lambda$ is a group in $\mathbb{R}^2$ such that for any $t_1,t_2\in \Lambda$, their summation is also $t_1+t_2 \in \Lambda$. This property makes the lattice codes the building blocks of \acrshort{cpnf} as for many simultaneous transmission techniques.

In a fundamental \acrshort{cpnf} network as given in Fig.~\ref{fig:cpnf}, $N$ nodes encode their message $w_n \in \{1,...,N\}$ to the lattices as 
\begin{equation}
    x_n = \mathcal{E}(w_n),
\end{equation}
where $\mathcal{E}$ is the encoding function. Then, $N$ nodes simultaneously transmit their message $x_n$ to the channel. A relay obtains the superimposed signal
\begin{equation}
    y = \sum_{n=1}^N h_n x_n + z,
\end{equation}
where $h_n$ is the channel coefficient and $z$ is the \acrshort{awgn}. The \acrshort{cpnf} decoder scales the received message as
\begin{equation}
    \alpha y = \sum_{n=1}^N \alpha h_n x_n + \alpha z ,
\end{equation}
and then quantizes the scaled signal to the closest lattice as follows
\begin{equation}
    \alpha y = \underbrace{\sum_{n=1}^N \beta_n x_n}_\text{approximation of $\alpha y$}  + \underbrace{\sum_{n=1}^N (\alpha h_n-\beta_n)x_n \beta z}_\text{effective noise}.  
\end{equation}

Here, the receiver uses $\alpha y$ to approximate the appropriate coefficients $\beta_n$ and obtains the following function of the codewords

\begin{equation}
    \sum_{n=1}^N \beta_n x_n. 
\end{equation}

A single relay in the given scenario only obtains an integer function of the messages which gives no information about the individual messages. However, the receiver can detect the messages if it obtains independent functions as many as the number of unknown messages. This idea inspired many studies in the literature and several aspects of the \acrshort{cpnf} have been investigated. Additionally, the \acrshort{cpnf} is extended to various channel and network models and several design challenges (e.g. \acrshort{csi} estimation) are addressed. We collect the related portion of these studies in Table~\ref{tab:cpnf} and classify them for their network topology, objective and performance metrics. 


\begin{table*}[t]  \caption{An overview of \acrshort{cpnf} studies.}
  \centering 
  \begin{tabular} { p{3.7cm}|p{1cm}|p{1.9cm}|p{5.5cm}|p{3.5cm}  }
  Study &  Year & Network & Contribution & Performance metric \\

  \hline \hline

Nazer and Gastpar~\cite{Nazer2011} & 2011  & $K \times K \times 1$ & \acrshort{cpnf} is proposed. & \multirow{2}{3.5cm}{Achievable rates} \\ \cline{1-4}

Nazer and Gastpar~\cite{Ning2011} & 2011  & Two-way  & Considered for \acrshort{plnc}. &  \\ \hline

Huang \textit{et al.}~\cite{Huang2013a} & 2013 & Multi-way & Rate optimization is investigated. & \multirow{7}{3.5cm}{Sum rates} \\ \cline{1-4}

Tan and Yuan~\cite{Tan2015, Tan2016} & 2015 & $K \times L \times 1$, multi-hop & \acrshort{cpcnf} is proposed. &  \\ \cline{1-4}

Nokleby and Nazer~\cite{Nokleby2013} & 2013 & $2 \times 2 \times 1$ & Amplify-and-Compute model proposed. &  \\ \cline{1-4}

Ntranos \textit{et al.}~\cite{Ntranos2013}, Tan \textit{et al.}~\cite{Tan2014} & 2013,4 & $K \times 1$, $K \times L \times 1$ & Asymmetric power allocation case is considered. & \\ \cline{1-4}

Pappi \textit{et al.}~\cite{Pappi2015} & 2015 & \acrshort{cran} &  Investigated from coalition game perspective. &  \\ \hline

El Soussi \textit{et al.}~\cite{ElSoussi2014} & 2014 & $2 \times 1 \times 1$ cooperative & Rate optimization is investigated. & \multirow{4}{3.5cm}{Symmetric rates} \\ \cline{1-4}

Ordentlich \textit{et al.}~\cite{Ordentlich2014} & 2014 & $K \times K$ &  Interference channel considered. &  \\ \cline{1-4}

Zhu and Gastpar~\cite{Zhu2015} & 2015 & Two-way &  Input distributions are investigated. &  \\ \hline

Wang \textit{et al.}~\cite{Wang2012} & 2012 & Multi-way & Outage prob. investigated. & Outage probability \\ \hline

Song \textit{et al.}~\cite{Song2011, Chen2013} & 2011,3 & $2 \times 2 \times 1$, $3 \times 3 \times 1$ & Inverse \acrshort{cpnf} model proposed. & \multirow{6}{3.5cm}{Rate region} \\ \cline{1-4}

Huang \textit{et al.}~\cite{Huang2013} & 2013 & Two-way & Capacity bounds investigated. &  \\ \cline{1-4}

Zhu and Gastpar~\cite{Zhu2013} & 2013 & $K \times 1$ & Considered for cognitive radios. &  \\ \cline{1-4}

Nazer and Gastpar~\cite{Nazer2014} & 2014 & $K \times 1$ &  Adapted to DMC. &  \\ \cline{1-4}

Lim \textit{et al.}~\cite{Lim2016, Lim2017, Lim2018, Lim2019} & 2016-9 & $K \times L \times 1$ &  Joint typicality decoder proposed. &  \\ \hline
%

Pappi \textit{et al.}~\cite{Pappi2013} & 2013 & $K \times L \times 1$ & \acrshort{cee} investigated. & \multirow{3}{3.5cm}{Computation rate} \\ \cline{1-4}

Ordentlich \textit{et al.}~\cite{Ordentlich2015} & 2015 & $2 \times 1$ &  Feedback included. &  \\ \cline{1-4}

Hong and Caire~\cite{Hong2011, Hong2012, Hong2013} & 2011-3 & $K \times L$ & A low complexity design presented. &  \\ \hline

Liu~\cite{Liu2014} & 2014 & Two-way & Channel inversion precoding considered. & Achievable rates, \acrshort{ser} \\  \hline

Tunali \textit{et al.}~\cite{Tunali2012, Tunali2015} & 2012,5 & $K \times L \times 1$ & Eisenstein integer lattices considered. & Outage prob., \acrshort{ser} \\ \hline

Wang and Burr~\cite{Wang2014} & 2014 & $2 \times 1 \times 2$ &  Coding gain improved with \acrshort{ldlc}. & \acrshort{ser} \\ \hline

Mejri \textit{et al.}~\cite{Mejri2012, Mejri2015} & 2012,5 & $K \times 1 \times 1$, $K \times 1$   & Decoding schemes compared. & Error probability \\ \hline

Wei and Chen ~\cite{Wei2012a} & 2012 & Two-way & Fincke-Pohst code search implemented & Average rate, zero entry prob.\\ \hline

Niesen and Whiting~\cite{Niesen2012} & 2012 & $2 \times 2 \times 1$ & Degrees of freedom investigated. & Degrees of freedom \\ \hline

Feng \textit{et al.}~\cite{Feng2012, Feng2016} & 2013 & $K \times 1$ & Blind \acrshort{cpnf} (without \acrshort{csi}) presented. & Throughput, complexity \\ \hline

Najafi \textit{et al.}~\cite{Najafi2013} & 2013 & $K \times L \times 1$ &  Synchronisation problems investigated. & Outage rate, average rate \\ \hline

Sakzad \textit{et al.}~\cite{Sakzad2014} & 2014 & $K \times L \times 1$ &  Phase precoding included. & Equation error rate \\ \hline

Wen \textit{et al.}~\cite{Wen2015} & 2015 & $K \times L \times 1$ &  \acrshort{svp} considered with sphere decoding. & Average computation rate \\ \hline

Nokleby and AAzhang~\cite{Nokleby2016} & 2016 & $K \times L \times 1$ &  Node cooperation case investigated. & Computation rate, outage prob. \\ \hline

Goldenbaum \textit{et al.}~\cite{Goldenbaum2016} & 2016 & $2 \times 2 \times 2$ &  Designed for \acrshort{ofdm} and 5G. & Message rate \\ \hline

Zhu and Gastpar~\cite{Zhu2017} & 2016 & $2 \times 1$ &  Typical sumsets of lattices investigated. & Density of the sets \\ \hline

Huang and Burr~\cite{Huang2016, Huang2017a, Huang2017} & 2016,7 & $K \times L \times 1$, $K \times 1 \times 1$ & A low complexity coefficient selection design suggested. & Cumulative distribution func. \\ \hline

Goseling \textit{et al.}~\cite{Goseling2013, Goseling2014} & 2013,4 & $K \times L$ &  Random access included. & Throughput \\ 

  \end{tabular}

  \label{tab:cpnf}
\end{table*}

In~\cite{Ning2011}, Nazer and Gastpar propose a \acrshort{plnc} scheme based on the nested lattice codes and \acrshort{cpnf}. It has been shown that the lattice codes can be exploited to transfer a function of the inputs to a sink node and the receiver can recover the messages if it obtains enough functions. The study considers a two-way relay channel and provides an introduction to the existing \acrshort{plnc} approaches. Later, the study proposes the lattice code based \acrshort{plnc} and compares with other schemes considering transfer rates.

In~\cite{Song2011}, a complementary scenario to the \acrshort{cpnf} scheme, inverse compute-and-forward, is considered. The \acrshort{cpnf} scheme computes a function of transmitted messages at relay nodes. The proposed network aims to recover back the computed functions of the \acrshort{cpnf} at the receiver. In the proposed network model, two \acrshort{cpnf} relays send their computed functions to a sink node. The inverse \acrshort{cpnf} is considered as a cascade to the traditional scheme and the rate region of the cascade network is investigated. It is shown that the proposed cascade network outperforms the traditional pairwise communication based relay networks on the rate region. In~\cite{Chen2013}, the authors extend their previous inverse \acrshort{cpnf} study to three transmitters scenario and investigate the rate region. Their results show that transmitting equations with a correlation between them provides superior performance than transmitting independent equations.  

Challenges of \acrshort{cpnf} on lattice decoding is considered in~\cite{Mejri2012}. The study analyzes the lattice decoders that are suitable to practical scenarios. Specifically, the performance of the maximum likelihood decoder, Diophantine approximation and the sphere decoder is investigated. The computer simulations are used to compare the decoders and to verify the previous theoretical results. Also, the performance of one and two dimensional lattice codes are investigated and it is shown via simulations that the performance degrades for larger constellations. This study is later extended to provide an overall base on the decoding of the \acrshort{cpnf} in~\cite{Mejri2015}. In addition to the previous study,~\cite{Mejri2015} includes a novel maximum a posteriori (\acrshort{map}) decoder and a Diophantine approximation based maximum likelihood decoder.

A two-way relay channel with \acrshort{cpnf} is considered in~\cite{Wei2012a} and~\cite{Liu2014}. The study~\cite{Wei2012a} proposes a Fincke-Pohst strategy based code search algorithm to find the appropriate coefficients. In~\cite{Wei2012}, the authors consider a multi-source multi-relay network to maximize the network flow with \acrshort{cpnf}. Contrary to the traditional \acrshort{cpnf} which optimizes the network coefficients separately for each relay, the proposed design jointly optimizes the coefficient matrix for all relays. The method utilizes a candidate set search algorithm based on the Fincke-Pohst strategy (as in~\cite{Wei2012a}) to select the coefficients. The performance of the proposed method is investigated with simulations. In~\cite{Liu2014}, a channel inversion precoding is proposed. The achievable rates and the symbol error rate (\acrshort{ser}) of the channel inversion precoded \acrshort{cpnf} are calculated and it is stated that the proposed precoding improves the performance of the \acrshort{cpnf}.

The communication between the nodes of a hexagonal lattice network is analyzed in~\cite{Goseling2012}. Four communication models are derived depending on the broadcast and superposition communications between the nodes. Two models that are applicable with \acrshort{cpnf} are investigated on the subject of network capacity. The study achieves an improved lower bound compared to the previous studies. The results reveal that the minimum transport capacity of the broadcast or superposition enabled case ($3/7$) is larger than the maximum capacity of the disabled cases ($2/5$).

In~\cite{Tunali2012} and~\cite{Tunali2015}, the alphabet of the lattice codes that is used in \acrshort{cpnf} is restricted to the Eisenstein integers\footnote{Eisenstein integers are the complex numbers in the form of $z= a+ b \exp[2\pi i/3]$ where $a,b \in \mathcal{Z}$}. The traditional \acrshort{cpnf} (as in~\cite{Nazer2011}) uses integer-based lattice codes and the study exploits the Eisenstein integers to obtain a better pair of nested lattice structure (the coarse and the fine lattice). It is shown that the outage performance and the error-correction performance of the proposed codebooks are superior to the integer-based lattice codebooks. 

In traditional \acrshort{cpnf}, received signals at the relays are scaled up in order to ensure that the coefficients are close to an integer. This is a result of the lattice codes that involve only integer codebooks and scaling the signals up also amplifies the noise at the receiver~\cite{Nazer2011}. Eventually, scaling the signals up establishes a Diophantine trade-off between the amplified noise level and the approximation performance. The Diophantine trade-off of the \acrshort{cpnf} scheme is investigated in~\cite{Niesen2012} and it is stated that the asymptotic rate of the scheme in~\cite{Nazer2011} is below the \acrshort{mimo} schemes. The authors design a novel compute-and-forward model that benefits from the interference alignment (\acrshort{ia}) in~\cite{Niesen2012}. The proposed model is shown to reach the same degrees of freedom with the \acrshort{mimo} scheme. 

In~\cite{Tan2015} and~\cite{Tan2016}, the compute-compress-and-forward (\acrshort{cpcnf}) method is proposed to establish an efficient multi-hop design of the \acrshort{cpnf}. The main idea behind the \acrshort{cpcnf} is the fact that relay forwarding rates can exceed the information rate of the sources. For this reason, \acrshort{cpcnf} includes a compressing phase to improve network efficiency (e.g power gain). The compressing and the recovering algorithms of the \acrshort{cpcnf} is designed and verified with numerical results. Later, the authors generalize the \acrshort{cpcnf} method in which the compression algorithm includes more operations and shows better compression performance as demonstrated with simulations. The same problem that results from the redundant forward rate is also considered in~\cite{Aguerri2016}. However, in~\cite{Aguerri2016}, the compression is applied at the symbol level rather than the message level and introduces a mapping to the system.

In~\cite{ElSoussi2014}, a cooperative relay network is considered, where two nodes are able to send their messages to a receiver both directly and over a relay. The study investigates two coding methods (\acrshort{cpnf} and \acrshort{cnf}) that are based on lattice codes and aims to optimize the symmetric rate. The authors propose an iterative coordinate descent method that focuses on the power allocation and integer coefficient selection processes for the optimization problem. The results reveal that \acrshort{cpnf} shows better performance than lattice-based \acrshort{cnf}. This work is later extended to a multi-user multi-relay cooperative scenario in~\cite{Soussi2014}. Cooperative networks are also considered in~\cite{Tseng2014,Jeon2016, Jlassi2016, Hasan2017, Jlassi2018}. The authors in~\cite{Jeon2016, Jlassi2016, Jlassi2018} propose an \acrshort{cpnf} method for the two transmitter single relay two receiver networks. The \acrshort{cpnf} is considered for the same scenario with single receiver in~\cite{Tseng2014, Hasan2017}.

The pairwise \acrshort{cpnf} model is applied to multi-way relay channels (\acrshort{mwrc}s) in~\cite{Huang2013a}. The pairwise structure of the network is accomplished with two phases: the broadcast transmission phase and the multiple access transmission phase. The sum rates of the pairwise \acrshort{cpnf} case and the pairwise successive transmission case are derived for the \acrshort{mwrc} and compared with each other. 

The outage probability of the \acrshort{cpnf} scheme is derived in~\cite{Wang2012} for \acrshort{mwrc}s. Also the \acrshort{cpnf} is compared with the non-network coding scheme with respect to the outage probabilities. The results show that in a canonical two-way relay channel, \acrshort{cpnf} achieves $7$ dB gain against the non-network coding at the outage probability of $10^{-2}$.

In~\cite{Nokleby2013}, amplify-and-compute method is proposed, which combines \acrshort{cpnf} and \acrshort{anf} methods. The relays in the network receive the superpositioned lattice codes from the sources as in \acrshort{cpnf} and transmit to the next network layer as in \acrshort{anf}. 

A \acrshort{cpnf} scheme that allows asymmetric power allocation to the nodes is proposed in~\cite{Ntranos2013} and~\cite{Tan2014}. The method in~\cite{Ntranos2013} is based on the lattice codes in which a fine lattice and a coarse lattice provide codebooks that are decodable and under the power limit respectively. The method maps the messages to the codebooks depending on the power and noise tolerance. Specifically, the top of the message vector is set to zero depending on the power of the codebook and the bottom of the message vector is set to zero depending on the noise tolerance.

A bi-directional relay network in which two nodes exchange information through a relay node under an inter-symbol interference channel is considered in~\cite{Huang2013}. The proposed model is separated into a multiple access phase and a broadcast phase and the capacity region is derived. The inner bound of the capacity region is computed with the help of the \acrshort{cpnf} method and the outer bound is computed with the cut-set argument given in~\cite{Cover2005}. The numerical results revealed that the proposed \acrshort{cpnf} scheme has a higher exchange rate than the \acrshort{dnf}.

The complexity reduced version of the \acrshort{cpnf} is proposed in~\cite{Hong2011}, which only involves scaling, offset and scalar quantization at the receivers. The method aims to reach the same capacity of the \acrshort{cpnf} with the low-complexity, low-power decentralized antenna networks. For this purpose, the method considers quantization at the receivers as a part of the wireless channel. The numerical results show that the computation rate of the quantized \acrshort{cpnf} is within the shaping error of 0.25 bits per symbol compared to the traditional \acrshort{cpnf} of~\cite{Nazer2011}. The authors extended their work to downlink scenario of the quantized \acrshort{cpnf} in~\cite{Hong2012}. The study derives the computation rate of the proposed scheme and compares it with the downlink \acrshort{cpnf} via simulations. The reverse \acrshort{cpnf} is generalized and extended in~\cite{Hong2013} to cover additional scenarios and to include comprehensive simulation results and comparisons.

Another low-complexity \acrshort{cpnf} scheme is given in~\cite{Hejazi2013}. The outage probability of the proposed model is derived and compared with the standard \acrshort{cpnf} scheme. Additionally, channel estimation error (\acrshort{cee}) is introduced to the system and its effect is investigated. The results show that the proposed simple method is also more resistant to the \acrshort{cee} than the traditional \acrshort{cpnf}. In~\cite{Hejazi2016}, the authors extend their study and propose two \acrshort{cpnf} based methods. The first method is designed to reduce the computational complexity and the second method is proved to have better performance compared to the traditional \acrshort{cpnf}, which is verified via simulations. 

The complexity of the coefficient selection of the \acrshort{cpnf} is considered in~\cite{Sahraei2014, Huang2016, Huang2017a, Huang2017}. The authors propose a low-complexity algorithm to optimize the integer coefficients that is essential for the \acrshort{cpnf} performance in~\cite{Sahraei2014}. This work is later improved and generalized to cover \acrshort{cpnf} and integer-forcing algorithms in~\cite{Sahraei2017}. The same problem is studied in~\cite{Huang2016, Huang2017a}.  An exhaustive search algorithm and a lattice reduction algorithm is proposed in~\cite{Huang2016, Huang2017a} to reduce the hardness of the coefficient selection problem. Also in~\cite{Huang2017}, a low-complexity coefficient method is proposed for massive \acrshort{mimo} enabled \acrshort{cpnf} networks.

The effect of the channel estimation error to the performance of \acrshort{cpnf} scheme is analyzed in~\cite{Pappi2013}. The computation rate region for the imperfect channel estimation case is derived and the expression is closely approximated for the Gaussian distributed \acrshort{cee}. Additionally, the distribution of the rate loss is given in the closed-form. The simulations are used to demonstrate the vulnerability of the \acrshort{cpnf} to the \acrshort{cee}. 

The traditional \acrshort{cpnf} requires \acrshort{csi} to decide the appropriate scale factors which are essential in the decoding of the integer lattices. Otherwise, the non-integer channel coefficients increase symbol error. In~\cite{Feng2012, Feng2016}, a practical \acrshort{cpnf} scheme is proposed that does not require \acrshort{csi} to compute the most suitable scale factors. Instead, the proposed method chooses sub-optimal however sufficient scaling factors to gain from the system complexity. Simulations show that in some cases, the computation complexity can be reduced ten times when compared to the \acrshort{cpnf} of~\cite{Nazer2011}.  

The synchronization problem of the \acrshort{cpnf} is considered in~\cite{Najafi2013}. \acrshort{cpnf} networks can exhibit asynchronization of the nodes as a result of the decentralized node structure of the network. The study solves the symbol asynchronization problem with an equalizer by converting the nature of the network from asynchronous to synchronous. The frame asynchronization is solved by eliminating delays with multiple antennas at the relay node. Also, it is shown that the achievable rate can be maximized for all \acrshort{snr} regions by applying a linear filter.

A phase precoding method for the \acrshort{cpnf} scheme is proposed in~\cite{Sakzad2014} for multi-user multi-relay networks. The objective of the method is to reach higher computation rates than the traditional \acrshort{cpnf}. However, it also requires an optimal precoding matrix and an optimal network equation matrix to fulfill that objective. For this reason, the study introduces a partial feedback channel between the relays and the nodes since the precoding matrix is needed at the nodes and the network equation matrix has to be computed at the relays. The relays compute the optimal precoder and the network equations, then forward the precoder information to the nodes through the feedback channel. With the simulations, the study shows that the proposed phase precoding can improve the equation error rate.

Another feedback enabled \acrshort{cpnf} method is given in~\cite{Ordentlich2015}. The method aims to design the optimal \acrshort{cpnf} model to achieve the maximum computation rates for the scenario that transmitters have access to an ideal feedback channel towards the relay. The method is designed for two users (and a relay) networks and it is demonstrated that the proposed scheme obtains better computation rates than the \acrshort{cpnf} without feedback.


\acrshort{cpnf} is considered for the Gaussian multi-user interference channels in~\cite{Ordentlich2014}. A comprehensive study is given on the approximate sum capacity and the capacity bounds. In~\cite{Nazer2014}, \acrshort{cpnf} is investigated for the discrete memoryless channels. The lattice codes are considered for the complex modulo arithmetics in~\cite{Vazquez-Castro2014}. It is shown that only five lattice code families are capable of complex modulo arithmetics over the Euclidean geometry and their coding gains are calculated. In~\cite{Pappi2015}, \acrshort{cpnf} scheme is considered for the cloud-radio access networks (\acrshort{cran}s). The study aims to maximize the information flow from the nodes to the \acrshort{fc} of the network. For this reason, a coalition game is designed that maximizes the defined profits. Exploiting the interference of signals is considered for the fifth generation ($5$G) networks in~\cite{Goldenbaum2016}. The objective is to provide channel access to a massive number of nodes that are required by the \acrshort{iot}. For this purpose, the study combines the \acrshort{plnc} with the pulse shaped orthogonal frequency division multiplexing (\acrshort{ofdm}).

In~\cite{Wang2014}, a low density lattice codes (\acrshort{ldlc}) based \acrshort{cpnf} method is proposed in order to reach high coding gains. In~\cite{Wen2015}, the authors propose a sphere decoding method to maximize the computation rate by considering the problem in hand as a shortest vector problem (\acrshort{svp}). The sumsets can be defined in simple terms as the set of received lattice points which is the sum of the transmitted lattice points. In~\cite{Zhu2017}, typical sumsets are defined and analyzed according to their sizes, distributions and densities. The study aims to obtain results that can improve the performance of the lattice decoding in \acrshort{cpnf}. 

Cooperation between the transmitters is considered in~\cite{Nokleby2016}. This is different than the cooperative networks as in Fig.~\ref{fig:cooperative}. In this study, cooperation indicates that the nodes can partially hear the messages of the other nodes which resembles the diversity improvement of a multiple antenna network. The results reveal that the partial cooperation between the nodes can increase the computation rate almost to the capacity.

In~\cite{Ordentlich2012}, a \acrshort{cpnf} transform is proposed in which the \acrshort{wmac} is transformed to a modulo-lattice \acrshort{mimo} channel with the help of \acrshort{sic}. Joint typicality decoders are adopted to the \acrshort{cpnf} method in~\cite{Lim2016, Lim2017, Lim2018, Lim2019}. The \acrshort{cpnf} is also considered for random access channels in~\cite{Goseling2013, Goseling2014, Ashrafi2018}. The impact of the input distribution to the computation rate of the \acrshort{cpnf} is considered in~\cite{Zhu2015} for the Gaussian \acrshort{wmac}. It is shown that the Gaussian input distribution is not optimal and the computation rates can be improved if the input distributions are chosen wisely. In~\cite{Zhu2013}, \acrshort{cpnf} is extended to cognitive radio networks.

In conclusion, PLNC and \acrshort{cpnf} methods aims to improve the total throughput of a network by exploiting superposition of signals. Especially \acrshort{cpnf} is widely investigated for various network models in the literature and essential theoretical results are obtained. Major drawback of these network coding schemes come from the perfect CSI requirement, perfect synchronization requirement and computational burden. Literature also includes studies that investigate and partially overcome these problems. Lastly, the literature lacks testbed implementations. We believe that the implementation of \acrshort{cpnf} schemes is an important future direction. In the following section, we present the computation and function alignment methods that are inspired by the function alignment and \acrshort{cpnf} methods.








%% file: compalignment.tex
\section{Interference / Computation / Function Alignment}

The interference alignment (\acrshort{ia}) is an interference management technique and can be compared with the multiple access methods for their application purpose. The conventional multiple access methods divide the time and frequency resources among the users. In the \acrshort{ia}, all users share the same resources, however, the \acrshort{ia} algorithm affects the transmitted signals (precoding) such that the received signals are aligned into two subspaces. As the algorithm aims, the unintended signals (the interference from the other users) fall under one subspace and the intended signal can be extracted from the other subspace. We are partially interested in the \acrshort{ia} since, on one hand, the network model enables the interference; on the other hand, it aims to cancel the interference instead of benefiting from it. 

We are much more interested in the computation alignment method that is inspired by the \acrshort{ia} and the \acrshort{cpnf}. Similar to the \acrshort{ia}, the computation alignment divides the signals into subspaces, however, the aligned signals are not discarded, instead, the interference is exploited for the computation. In this section, we present an elementary description of the \acrshort{ia} and refer the readers to~\cite{Zhao2016} and~\cite{Ayach2013} for detailed information. Later, we continue with the computation alignment and present the current studies. The fundamental \acrshort{ia}, computation and function alignment studies are exhibited in Table~\ref{tab:ia}.

\begin{table*}[t]  \caption{An overview of \acrshort{ia}, computation alignment and function alignment studies.}
  \centering
\begin{tabular}{ p{3cm}|p{1.3cm}|p{1.3cm}|p{7cm}|p{1.6cm}  }
  Author  &  Year  & Network model & Contribution & Performance metric  \\

  \hline \hline

Niesen \textit{et al.}~\cite{Niesen2011,Niesen2013} & 2011,2013  & \multirow{5}{2cm}{$K \times K $}   & Provides a capacity approximation for multi-layer networks that is independent of network depth. & Capacity approximation \\  \cline{1-2} \cline{4-5}

Goela \textit{et al.}~\cite{Goela2012} & 2012  &  & Investigates coding schemes that reach computation capacity with network decomposition.  & Coding capacity \\ \hline

Suh \textit{et al.}~\cite{Suh2012, Suh2016} & 2012, 2016  & \multirow{5}{2cm}{$2 \times 2 $}  & Derives a new upper bound on the computing capacity and propose a network decomposition theorem. & \multirow{3}{2cm}{Computing capacity} \\ \cline{1-2} \cline{4-4}

Suh and Gastpar~\cite{Suh2013a} & 2013  &  & Considers feedback for function alignment. &  \\ \cline{1-2} \cline{4-5}

Suh and Gastpar~\cite{Suh2013} & 2013  & & Investigates the scenarios where network decomposition is optimal.  &  Symmetric capacity \\

  \end{tabular}

  \label{tab:ia}
\end{table*}

\begin{figure}[t]
\begin{subfigure}[t]{.48\textwidth}
  \centering
  \includegraphics[width=.77\linewidth]{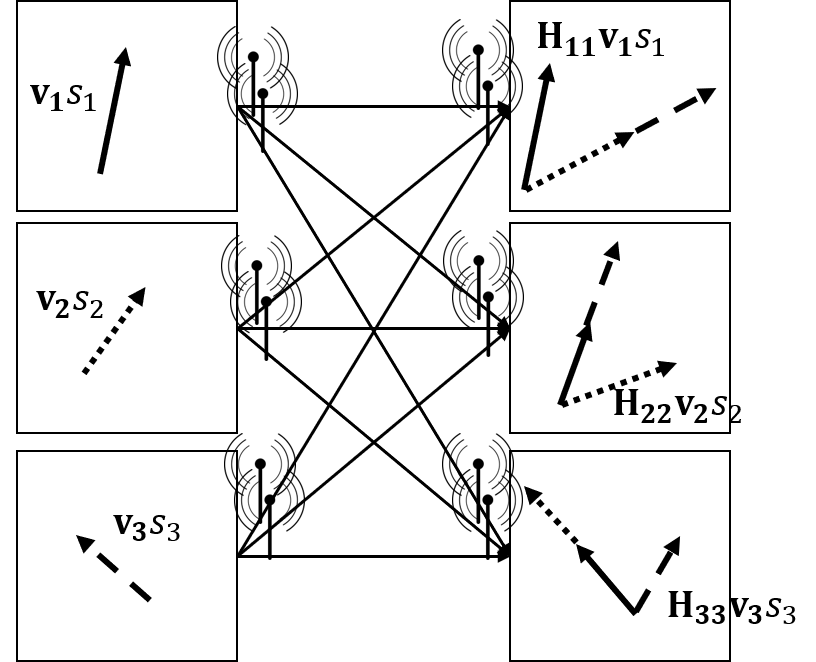}  
  \caption{}
  \label{fig:intalign} \vspace*{.9cm}
\end{subfigure} 
\begin{subfigure}[t]{.48\textwidth}  \hspace*{.6cm}
  \centering \vspace*{.5cm}
  \includegraphics[width=.8\linewidth]{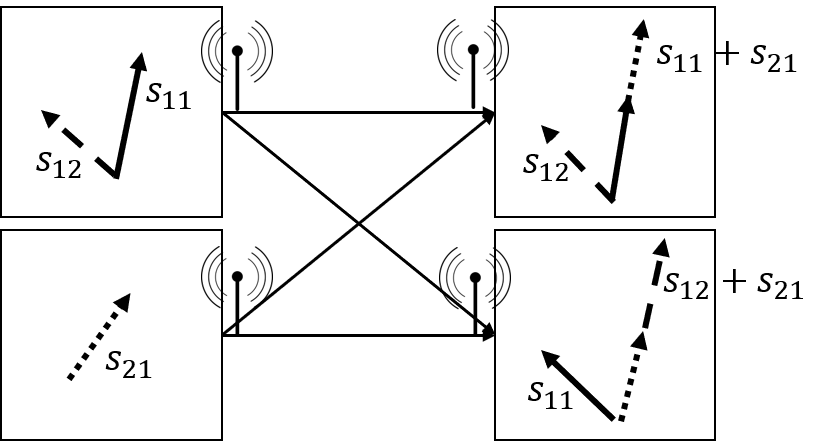}  
  \caption{}
  \label{fig:compalign}
\end{subfigure}
\caption{Vector illustration of the interference alignment and computation alignment methods. (a) Interference alignment (b) Computation alignment}
\label{fig:alignment}
\end{figure}

The \acrshort{ia} studies are mainly centered upon the space, frequency, or time dimensions to align the interference~\cite{Zhao2016}. We focus on a simple space dimension example, which is based on multiple antenna techniques. Consider a  \acrshort{mimo} network that consists of three transmitters and three receivers all of which equipped with two antennas as shown in Fig.~\ref{fig:intalign}. After the simultaneous transmission, the first receiver obtains the following signal.
\begin{equation}\label{eq:alignment}
    \mathbf{y_1} = \mathbf{H_{11}} \mathbf{v_1} s_1 + \mathbf{H_{21}} \mathbf{v_2} s_2 + \mathbf{H_{31}} \mathbf{v_3} s_3 + \mathbf{n_1}.
\end{equation}

The subscripts indicate the users, i.e. $\mathbf{y_j}\in \mathbb{C}^{2 \times 1}$ is the received signal vector of the $j^{th}$ receiver. The first and second row of the vector are the received signal at the first and second antenna, respectively. $\mathbf{H_{ij}} \in \mathbb{C}^{1 \times 2}$ is the channel matrix vector from the $i^{th}$ transmitter to the $j^{th}$ receiver. $\mathbf{v_i} \in \mathbb{C}^{2 \times 2}$ is the diagonal precoding matrix of the $i^{th}$ transmitter, $s_i$ is the message of the $i^{th}$ transmitter and the $\mathbf{n_j} \in \mathbb{C}^{2 \times 1}$ is the Gaussian noise vector at the $j^{th}$ receiver, where its rows are the noise of the first and second antenna, respectively.

Assuming perfect \acrshort{csi} at the receivers, the \acrshort{mimo} algorithms require three antennas since each of the three users' inputs is an unknown variable. In \acrshort{ia}, the precoding vector ($\mathbf{v_i}$) elegantly aligns the two unknown vectors (the interference from the other users) such that \eqref{eq:alignment} can be written with two unknown vectors. After that, the two unknown variables can be solved with the two equations. The vector representation of how the \acrshort{ia} works is illustrated in Fig.~\ref{fig:alignment}. Each node is represented with a square and the transmitted and received signals are given in the inside of these squares. The precoding at the transmitter divides the three vectors into two vectors; one of them is the intended vector and the other is the aligned (unintended) signals.

The perspective given above exploits the physical layer to suppress the interference. The computation alignment method is inspired from this perspective, however, it focuses on the unintended aligned vector for the function computation, which genuinely exploits the signal interference. A simple two transmitter two receiver network example of the computation alignment is given in Fig.~\ref{fig:compalign}. In this example, the time dimension is used to create the subspaces. Assume that the first transmitter aims to send $s_{11}$ and $s_{12}$ while the second transmitter aims to send $s_{21}$. In this scenario, transmit vectors adjust the messages as follows
\begin{equation}
\begin{split}
    \begin{bmatrix}
x_1(t_1)\\
x_1(t_2)
\end{bmatrix}&
= \mathbf{v_{11}} s_{11} + \mathbf{v_{12}} s_{12} \\
    \begin{bmatrix}
x_2(t_1)\\
x_2(t_2)
\end{bmatrix}&
= \mathbf{v_{21}} s_{21}, 
\end{split}
\end{equation}
where $t_1$, $t_2$ denote the time slots and $x_1$, $x_2$ denote the transmitted signals of the first and second transmitters, respectively. Proper selection of the transmit vectors lead to the following received signal vectors

\begin{equation}\label{eq:compalignment}
\begin{split}
    \mathbf{y_1} &= \begin{bmatrix}
1\\
1
\end{bmatrix} (s_{11} + s_{21}) + h\begin{bmatrix}
1\\
-1
\end{bmatrix} s_{21} + \mathbf{n_1} \\
    \mathbf{y_2} &= h\begin{bmatrix}
1\\
1
\end{bmatrix} (s_{12} + s_{21}) + \begin{bmatrix}
1\\
-1
\end{bmatrix} s_{11} + \mathbf{n_1},
\end{split}
\end{equation}
where the subscripts indicate the users, i.e. $\mathbf{y_j}\in \mathbb{C}^{2 \times 1}$ is the received signal vector of the $j^{th}$ receiver. However, subspaces are obtained with multiple transmissions rather than multiple antennas in this example, hence the first and second row of the vector are the received signal at the first and second time slots, respectively. After obtaining signals from two time slots, the first receiver can obtain $s_{11} + s_{21}$ by using $y_1(t_1)+y_1(t_2)$ and the second user can obtain $s_{11}$ by using $y_2(t_1)-y_2(t_2)$.

The design of the transmit vector enables computation alignment. A vectorial illustration of \eqref{eq:compalignment} is given in Fig.~\ref{fig:compalign}. It should be noted that the illustration omits the representation of the transmit vectors and the channel gain vectors in the figure for better appearance. As a result of the computation alignment, the receivers obtain the summation of the messages in the aligned subspace. Specifically, the first receiver obtains $s_{11}+s_{21}$ aligned and $s_{12}$ in the other subspace while the receiver two obtains $s_{12}+s_{21}$ and $s_{11}$.

In a relay network, \acrshort{cnf} method adds additional noise to the network in each layer. As a result, the approximation gap of the network capacity widens for the increasing number of network layers. The computation alignment scheme in~\cite{Niesen2013} and~\cite{Niesen2011} presents a relay network with an approximation gap that is independent of the layer depth. The computation alignment technique depends on lattice codes and the \acrshort{cpnf} to be able to recover the integer-valued messages. However, the \acrshort{cpnf} scheme also produce errors as a result of the non-integer channel gains. This problem is solved by \acrshort{ia} by dividing the channel into multiple subchannels and aligning. The results reveal that the approximation gap is not constant as opposed to \acrshort{cnf} results, it depends on fading characteristics. 

In~\cite{Goela2012}, multiple transmitter multiple receiver summation networks are considered. The scalar and vectorial linear codes are investigated for these networks and it is stated that the computation alignment is essential to reach the computation capacity.  For this purpose, the network is decomposed into sub-networks with the network equivalence theorems. Also, the linear coding capacity of the computation is derived for various channel parameters. 

In~\cite{Suh2012} and~\cite{Suh2016}, $2 \times 2$ modulo-2 sum networks\footnote{In particular Avestimehr-Diggavi-Tse (ADT) network is considered.} are considered. The study is inspired by the \acrshort{ia} and similar to the computation alignment given in~\cite{Niesen2011}, and named as the function alignment. A new upper bound is derived for the computing capacity of the two receiver \acrshort{cpnf} networks with linear codes in~\cite{Suh2012, Suh2016}. Also, the studies define a network decomposition theorem to divide the network into elementary subnetworks. Using the theorem, the computing capacity is generalized for the $N$-transmitter $N$-receiver \acrshort{cpnf} networks. In~\cite{Suh2013a}, the authors extend their previous work,~\cite{Suh2012}, to include feedback. The study derives the feedback included computing capacity and compare it with the no-feedback scheme (as in~\cite{Suh2012}). It should be stated that network decomposition is crucial to create subnetworks. The authors also investigate the network decomposition and its importance in \acrshort{cpnf} networks thoroughly in~\cite{Suh2013}.


In~\cite{Bross2016}, memoryless bivariate Gaussian sources are considered for a two-source one receiver network. The receiver aims to obtain the information of the two sources with the minimum distortion. In the paper, perfect causal feedback is assumed and the power-distortion relationship is investigated. The results are given as a function of the source correlation and \acrshort{snr}. Also, the necessary and sufficient conditions to reach the minimum distortion levels are derived. The study is later extended to two transmitter two receiver networks in~\cite{Song2015}.

Computation alignment is a form of interference alignment for the purpose of function computation. Although the next section will be purely devoted to the function computation, computation alignment differentiate from other studies by its working mechanism. In conclusion, IA allows interference of signals and removes the interference at the receiver. On the other hand, computation alignment exploits superposition to perform mathematical tasks over the air. IA has gained small attention in the literature and lacks testbed implementations. We expect that mainly the perfect CSI requirement prevents researchers from exhibiting a proof of concept. Simultaneous transmission is also the basis of function computation techniques that are inspired by lattice codes and \acrshort{cpnf}. Moreover, the function computation methods are considered for analog signals and the resulting analog function computation studies gained popularity in the literature.


%% file: functioncomputation.tex
\section{Function Computation}

Function computation is one of the most striking applications of simultaneous transmission. The joint source-channel coding paradigm by Gastpar and Vetterli~\cite{Gastpar2003} and later Nazer and Gastpar~\cite{Nazer2007} establish the basis for the function computation. This approach inspires a wide range of studies that often targets one of the two main aspects; computation or multiple access. The motivation behind the studies that focus on the multiple access aspect is generally to improve the network throughput as in \acrshort{cpnf} and considered in Section~\ref{sec: networkcoding}. Here, we present the studies that focus on the computation aspect as the digital function computation. The computation aspect later inspires the analog transmission based studies that purely focus on function computation. The motivation behind these studies is strictly computation related and presented as the analog function computation~\cite{Goldenbaum2013}.

\subsection{Digital Function Computation}


A list of the digital function computation studies is presented in Table~\ref{tab:dfc}. Nazer and Gastpar proposed the computation codes in~\cite{Nazer2007} that is based on lattices. The objective of the study is to send a linear function of multiple users to a receiver with the simultaneous transmission. The study investigates the achievable rates with the proposed computation codes and compares them with the separation based methods. Computation of linear functions is also considered in~\cite{Soundararajan2012} for a wireless network that consists of two correlated Gaussian sources. The work aims to find the optimum coding scheme that upper bounds the distortion at the received signal. The numerical results are given for the subtraction ($a-b$) and weighted addition ($a+2b$) functions as a function of the correlation coefficient of the two sources.

\begin{table*}[t]  \caption{An overview of digital function computation studies.}
  \centering
  \begin{tabular}{ p{4.2cm}|p{1.3cm}|p{1.3cm}|p{7cm}|p{1.6cm}  }
  Author  &  Year  & Network model & Contribution & Performance metric  \\

  \hline \hline

Jeon \textit{et al.}~\cite{Jeon2013, Jeon2014} & 2013, 2014  &  \multirow{5}{2cm}{$K \times L \times 1 $}  & Extends the lattice code computable function set by considering orthogonal components and derives an approximation of computation capacity. & Computation rate \\ \cline{1-2} \cline{4-5}

Wu \textit{et al.}~\cite{Wu2016, Zhang2016} & 2015, 2016  &   & Proposes a low-bandwidth, low-energy \acrshort{sdr} network architecture, \acrshort{stac}. & \acrshort{ser}, session rate \\ \hline

Nazer and Gastper~\cite{Nazer2007} & 2007  &  \multirow{10}{2cm}{ $K \times 1 $} & Proposes the computation codes and the \acrshort{comac} which enhance the communication performance by utilizing lattice codes and the joint source-channel coding.  & \multirow{10}{2cm}{Computation rate} \\ \cline{1-2} \cline{4-4}

Goldenbaum \textit{et al.}~\cite{Goldenbaum2013a, Goldenbaum2015a} & 2013, 2015  &  & Improves the reliability of nomographic function computation by adapting lattice codes to \acrshort{afc} based consensus methods. &  \\ \cline{1-2} \cline{4-4}

Jeon and Jung~\cite{Jeon2015, Jeon2016a} & 2015, 2016  &  & Provides non-vanishing computation rates (asymptotically positive) by allowing only a subset of nodes with high gains to transmit. &  \\ \cline{1-2} \cline{4-4}

Wu \textit{et al.}~\cite{Wu2019, Wu2020, Wu2019a, Wu2019b} & 2019  &  & Considers the wide-band implementation of \acrshort{comac} by allocating sub-functions to subcarriers. &  \\ \hline

Soundararajan and Vishwanath~\cite{Soundararajan2012} & 2012  &  $2 \times 1 $ & Derives a lower bound on the distortion of \acrshort{comac} and considers correlated source scenario. & Distortion rate  \\ \hline

Zhan \textit{et al.}~\cite{Zhan2011a} & 2011  &  Butterfly & \multirow{4}{7cm}{Investigates the duality between computation and communication aspect of simultaneous transmission schemes.}    & Distortion level \\ \cline{1-3} \cline{5-5}

Zhu \textit{et al.}~\cite{Zhu2017b, Zhu2019} & 2017, 2019  &  $2 \times 2 $ &  & Capacity region \\  \hline

Chen \textit{et al.}~\cite{Chen2020} & 2020 & $K \times 1 $ & A low-complexity transceiver model is designed to maximize the achievable function rate. & Achievable function rate

  \end{tabular}

  \label{tab:dfc}
\end{table*}

Goldenbaum \textit{et al.} generalize the computation of nomographic functions with nested lattice codes in~\cite{Goldenbaum2013a}. The model is based on the fact that the nomographic functions can be written in the form of \textit{pre} and \textit{post} processing functions as studied in the analog function computation research. However, the authors implement a digital model to reduce the destructive effects of the noise. The study also examines the required number of channel use and the accuracy performance of the system. Lattice codes are used in~\cite{Goldenbaum2015a} for the computation of nomographic functions. The study thoroughly analyzes the relation between the lattice codes and the nomographic functions and derive the computation rate performances. One of their observations reveals that any continuous function can be computed over the channel. 


In~\cite{Jeon2013, Jeon2014}, the authors include orthogonal components to their coding scheme in a similar manner that the \acrshort{tbma} benefits from the orthogonal signals. The method is interested in calculating the arithmetic summation and the type functions. The type function computes the histogram of the transmitted signals as explained in \acrshort{tbma}. Then the resulting statistics can be used to obtain the mean, variance, maximum, minimum and median functions. The \acrshort{wmac} is firstly decomposed into multiple modulo sum subchannels with nested lattice codes and linear network codes. Then the linear Slepian–Wolf source coding is used to calculate the desired functions. It is shown that the joint source-channel coding provides better performance than the separate coding schemes in certain cases.

The simultaneous transmitting and air computing (\acrshort{stac}) method is proposed in~\cite{Wu2016} and~\cite{Zhang2016} to improve the function computation capability and the network efficiency of data center networks. The proposed method is based on a traditional function computation model that combines communication and computation; however, the model is also developed upon an enhanced software-defined network structure that provides side information to the nodes. Computer simulations demonstrate the spectrum and energy efficiency of \acrshort{stac} in data center networks.

Function computation problem is considered for the fading \acrshort{mac}s in~\cite{Jeon2015} and~\cite{Jeon2016a}. The main idea behind the study is that only the nodes with high channel gains participate in the in-network computation rather than the whole network. The study investigates the computation rates of the proposed model for the fading channels.

A duality between the function computation problem and the multicast problem is considered in~\cite{Zhan2011a}. The function computation problem is interested in receiving a function (e.g. summation) of the transmitted messages. In the multicast problem, the objective is to receive individual messages. After defining the duality relation for the deterministic networks, Gaussian \acrshort{mac}s are considered. The achievable distortion levels are derived for the summation of two Gaussian sources in these networks. The results revealed that there is a constant gap between the cut-set bound and the distortion of the summation function of the independent Gaussian sources.

It is obvious that the superposition of the signals destroys the individual information of the transmitted messages. This can be viewed as renouncing the capability of the nodes to (multiple) access to the receiver~\cite{Zhu2017b, Zhu2019}. However, computation codes benefit from this idea to improve the efficiency of the computations in the network. Here, access to the individual information is traded with improved computation efficiency and this can be highly beneficial if the network is only interested in a function of the transmitted data. In other words, computation codes present a duality between the multiple access and computation. This duality is investigated in~\cite{Zhu2017b} and~\cite{Zhu2019} to check the existence of the computation codes that also allow the individual access to the receiver. The investigation results indicate that efficient computation codes prevent the individual access to the receiver.

The wideband computation over multiple access channel (\acrshort{comac}) schemes that are adopted for the frequency selective channels is proposed in~\cite{Wu2019, Wu2020, Wu2019a, Wu2019b}. These studies rely on the \acrshort{noma} and \acrshort{ofdm} models to use wideband frequencies. A \acrshort{noma} assisted function computation network is proposed in~\cite{Wu2019} and~\cite{Wu2020}. As given in the previous sections, \acrshort{noma} is a multiple access method that is based on the superposition of signals. In~\cite{Wu2019}, functions to be computed are intentionally divided into sub-functions and these sub-functions are computed simultaneously under different \acrshort{noma} access slots. As a result, the computations can be made at the wideband frequencies where the fading is more challenging. The results reveal that the proposed \acrshort{noma} assisted approach achieves higher computation rates and prevents vanishing computation. Additionally, expressions for the diversity order of the computation rate is derived in~\cite{Wu2020}. In~\cite{Wu2019a} and~\cite{Wu2019b}, the sub-functions are allocated to the \acrshort{ofdm} carriers and an optimization problem is considered for the power allocation. In~\cite{Chen2020}, a transceiver model is designed for digital function computation that reduces the time-complexity. The authors derive the achievable function rates of the proposed model by considering the number of nodes, the maximum value of messages and the quantization error threshold.

\subsection{\acrfull{afc}}

\begin{figure}[t]\center
  \includegraphics[width=1\linewidth]{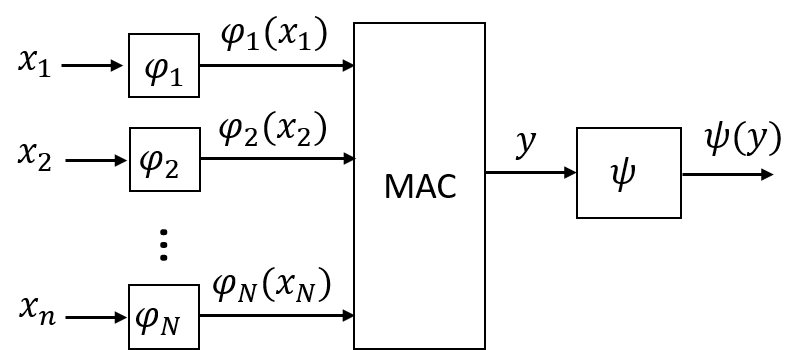}
  \caption{The network model of the \acrshort{afc} studies. $\psi$ and $\varphi$ denote \textit{pre}-processing and \textit{post}-processing functions, respectively.}
  \label{fig:afc}
\end{figure}

The main idea behin the \acrshort{afc} is to match the \acrshort{wmac} with the desired function. In its base form, the \acrshort{wmac} constitutes a natural summation operation with its superposition property. The \acrshort{afc} adjusts the channel with proper signal processing at the transmitter and receiver ends such that the \acrshort{wmac} can compute other mathematical operations. For this purpose, transmitters use \textit{pre}-processing functions, $\varphi_n (\cdot) : \mathbb{R} \rightarrow \mathbb{R}$, before transmitting their signals and the receiver applies a \textit{post}-processing function, $\psi (\cdot) : \mathbb{R} \rightarrow \mathbb{R}$, after receiving the superpositioned signal as depicted in Fig.~\ref{fig:afc}. The users simultaneously transmit their \textit{pre}-processed signals and the receiver obtains the following function output after the \textit{post}-process,
		\begin{equation} \label{nomographic}
			f(x_1,x_2,...,x_N) = \psi \left(\sum_{\substack{n=1}} ^{N}	\varphi_n(x_n)\right).
		\end{equation}

The functions that can be expressed as in \eqref{nomographic} with the summation operation are called nomographic functions. The \acrshort{afc} is firstly shown to be applicable to nomographic functions and later it is proven that any function can be computed with the \acrshort{afc}~\cite{Goldenbaum2013}. The \acrshort{afc} studies are listed in Table~\ref{tab:afc} with their contributions and performance metrics.

\begin{table*}[t]  \caption{An overview of analog function computation studies.}
  \centering
  \begin{tabular}{ m{4cm}|p{1.3cm}|p{8cm}|p{1.6cm}  }
  Author  &  Year  &  Contribution & Performance metric  \\

  \hline \hline

Goldenbaum \textit{et al.}~\cite{Goldenbaum2009a} & 2009  &  Computes functions (e.g. arithmetic mean) over the channel with low-complexity and low-energy consumption. & \multirow{10}{2cm}{Outage probability}  \\  \cline{1-3}

Goldenbaum and Sta\'{n}czak~\cite{Goldenbaum2010} & 2010  &   Extends~\cite{Goldenbaum2009a} to investigate the geometric mean. &  \\  \cline{1-3}

Sta\'{n}czak \textit{et al.}~\cite{Stanczak2012, Goldenbaum2013b} & 2012, 2013  &   Provides a complete theory of \acrshort{afc}. & \\  \cline{1-3}

Goldenbaum and Sta\'{n}czak~\cite{Goldenbaum2014} & 2014  &  Investigates the \acrshort{afc} for different \acrshort{csi} assumptions. &  \\  \cline{1-3}

Jeon and Jung~\cite{Jeon2018} & 2018  & Improves the traditional \acrshort{afc} against fading environment by utilizing causal \acrshort{csi}. &  \\  \cline{1-3}

Chen \textit{et al.}~\cite{Chen2018} & 2018   & Extends~\cite{Chen2018a}, considers the computation of multiple functions and investigates the method's performance. &  \\  \hline

Goldenbaum \textit{et al.}~\cite{Goldenbaum2013} & 2013   & Investigates and generalizes the functions that is computable with \acrshort{afc}. & \multirow{18}{2cm}{\acrshort{mse}} \\  \cline{1-3}

Limmer and Sta\'{n}czak~\cite{Limmer2014} & 2014   & Investigates the computation of $l_p$-norm functions with \acrshort{afc}. &  \\  \cline{1-3}

Huang \textit{et al.}~\cite{Huang2015} & 2015  &   Extends \acrshort{afc} to \acrshort{mimo} networks and includes imperfect \acrshort{csi}. &  \\  \cline{1-3}

Zhu \textit{et al.}~\cite{Zhu2018, Zhu2019a} & 2018   & Introduces a multi-modal AirComp technique with \acrshort{mimo} and beamforming that minimizes distortion. &  \\  \cline{1-3}

Farajzadeh \textit{et al.}~\cite{Farajzadeh2020} & 2020   & Removes the \acrshort{csi} requirement of AirComp. &  \\  \cline{1-3}

Wen \textit{et al.}~\cite{Wen2019} & 2019  &   Provides low-complexity and low-latency with \acrshort{mimo} AirComp. &  \\ \cline{1-3}

Ang \textit{et al.}~\cite{Ang2019a, Ang2019} & 2019  &   Reduce the complexity of the training process in massive \acrshort{csi} acquisition. &  \\\cline{1-3}

Li \textit{et al.}~\cite{Li2019a, Li2019} & 2018, 2019   & Considers an AirComp method that transfers power to the distributed nodes over the air to relax energy constraints. &  \\\cline{1-3}

Cao \textit{et al.}~\cite{Cao2019a} & 2019  &   Considers joint optimization of the transmit powers and the denoising factor to improve AirComp. &  \\\cline{1-3}

Basaran \textit{et al.}~\cite{Basaran2020} & 2020   & Provides an energy efficient AirComp method by exploiting the correlation between the measurements. &  \\\cline{1-3}

Chen \textit{et al.}~\cite{Chen2018a} & 2018  &   Considers non-uniform fading scenario in the proposed AirComp method and brings robustness with the uniform-forcing transceiver design. &  \\\hline

Limmer \textit{et al.}~\cite{Limmer2016} & 2015   & Provides a method that computes certain multivariate functions with nomographic function approximation. & Approximation error \\\hline

Goldenbaum \textit{et al.}~\cite{Goldenbaum2015} & 2015  &   Adapts computation codes to \acrshort{afc} to provide reliability and investigate its rates. & Computation rate \\\hline

Wang \textit{et al.}~\cite{Wang2015} & 2015  &   Without channel estimation, reduces the required number of channel uses on the computation of mean function with free deconvolution. & \multirow{4}{2cm}{Relative error } \\\cline{1-3}

Dong \textit{et al.}~\cite{Dong2020} & 2020  &  Uses Wirtinger flow to provide a low-complexity, low-latency AirComp method that requires no \acrshort{csi}). &  \\\hline

Chen \textit{et al.}~\cite{Chen2019} & 2019   & Investigates and compares the computation based and communication based methods. & Function rate \\\hline

Jakimovski \textit{et al.}~\cite{Jakimovski2011} & 2011   & \multirow{5}{8cm}{ Provides a testbed implementation. }  & Average error \\ \cline{1-2} \cline{4-4}

Sigg \textit{et al.}~\cite{Sigg2012} & 2012   &  & Mean error \\ \cline{1-2} \cline{4-4}

Kortke \textit{et al.}~\cite{Kortke2014a} & 2014  &    & Relative error \\ \cline{1-2} \cline{4-4}

Abari \textit{et al.}~\cite{Abari2015} & 2015  &    & \acrshort{cfo} \\ \cline{1-2} \cline{4-4}

Altun \textit{et al.}~\cite{Altun2018} & 2017  &   & \acrshort{mse} \\ \hline

  \end{tabular}

  \label{tab:afc}
\end{table*}

Goldenbaum and Sta\'{n}czak pioneered the studies that compute mathematical functions at the communication process. In~\cite{Goldenbaum2009a}, the communication system carries the information at the transmit powers which relieves the synchronization burden. Their method manages to compute a variety of functions which involve nonlinear functions. The authors lastly exhibit the error analysis of the arithmetic mean function. In~\cite{Goldenbaum2010}, Goldenbaum and Sta\'{n}czak  extend their previous work by analyzing the geometric mean function. They also compare their method with TDMA via simulations and present their results that \acrshort{afc} outperforms the TDMA on the function computation time. 

In their following works, Goldenbaum and Sta\'{n}czak provide an extensive theory of \acrshort{afc} in~\cite{Stanczak2012} and~\cite{Goldenbaum2013}. They analyze the function sets and network topologies that are compatible with \acrshort{afc}. Their results show that every function can be computed with \acrshort{afc} as the \textit{pre}-processing is independent of the computed function. 

The error performance of the estimators that are used in the receivers of \acrshort{afc} is analyzed in~\cite{Goldenbaum2013b} for arithmetic and geometric mean functions. Additionally, function computation simulations are performed for \acrshort{afc}, TDMA and CDMA models. The results showed that the \acrshort{afc} brings out better computation accuracy in less time than time or code divided networks.

Dependency of \acrshort{afc} to \acrshort{csi} is investigated in~\cite{Goldenbaum2014}. The letter firstly shows that the transmitters only need the magnitude of the \acrshort{csi} rather than the full \acrshort{csi}. Then it is proven that any \acrshort{csi} knowledge requirement on the transmitters can be removed if the receiver is equipped with multiple antennas. 

The computation of the $l_p$-norms with the \acrshort{afc} is investigated in~\cite{Limmer2014}. The authors later propose an algorithm to approximate certain continuous multivariate functions by the nomographic function in~\cite{Limmer2016}. In~\cite{Huang2016}, the max function which aims to obtain the maximum value at the \acrshort{fc} is implemented with the \acrshort{afc} and \acrshort{cdma}.

In~\cite{Goldenbaum2015}, Goldenbaum and Sta\'{n}czak investigate the relation between the reliability and efficiency of an \acrshort{afc} model from an information-theoretic point of view. In order to evaluate the relation in question, the computation rate metric is defined as the number of functions that can be computed per channel use. Then, achievable computation rates are given for linear combination and special polynomial functions. The letter states that the computation rate is dependant to the required accuracy level and the number of transmitters as well as the function type.

An \acrshort{afc} method based on the free deconvolution theorem (see~\cite{anderson_guionnet_zeitouni_2009} for the further information on the free deconvolution) is proposed in~\cite{Wang2015}. The contribution of the study is that the mean function can be computed with fewer channel uses and without the channel estimation.

In~\cite{Huang2015}, a \acrshort{mimo} \acrshort{afc} model is proposed that the \acrshort{fc} and the multiples transmitting nodes are equipped with multiple antennas. The method allows channel estimation errors in the system model to consider realistic scenarios and design a non-convex optimization problem to achieve the optimum solution in this scenario. The optimization problem that minimizes the worst-case mean square error (\acrshort{mse}) is converted to a simpler version and solved.

An adaptive \acrshort{afc} method is given in~\cite{Jeon2018}. Different from the traditional \acrshort{afc} methods, this model is based on the causal \acrshort{csi} at the transmitters. Simulation results of the study state that the proposed adaptive model shows better outage probability performance than the traditional \acrshort{afc}.

In~\cite{Zhu2018}, a function computation method that can compute functions for multiple variables is proposed. For example, multi-model sensor measurements such as temperature, humidity and pollution can be computed over the air simultaneously. This is possible with the utilization of beamforming technology over \acrshort{mimo} enabled nodes. The study designs an optimization problem that minimizes the sum mean-squared error at the receiver. The optimization problem is shown to be an NP-hard problem and an approximate version is solved with differential geometry. Also, the result of the optimization problem (based on Grassmann Manifold) is validated with computer simulations. This work is later extended to high-mobility sensing networks where an environment is monitored by unmanned aerial vehicles (\acrshort{uav}s)~\cite{Zhu2019a}. In addition to~\cite{Zhu2018}, the study includes enhanced equalization and channel feedback methods to provide reliable information exchange between the sensors and the receiver.  

Non-uniform fading in different nodes of an \acrshort{afc} network is a performance degrading problem. In order to mitigate the effects of non-uniform fading, a uniform-forcing transceiver design is proposed in~\cite{Chen2018a}. The authors formulate an optimization problem that minimizes the mean square error of the computed function output at the receiver. Their results show that a semidefinite relaxation is necessary to solve the problem and successive convex approximation can further increase the accuracy of the solution. The proposed transceiver model in~\cite{Chen2018a} is extended in~\cite{Chen2018} to compute multiple functions simultaneously. The nodes in the new network model are equipped with antenna arrays and zero-forcing beamforming technology. Beamforming technology is adopted to remove the interference of other functions that are computed simultaneously and multiple antenna arrays relieve the massive \acrshort{csi} knowledge requirement.  Also as in~\cite{Chen2018a}, the method is resilient to the non-uniform fading problem. The performance of the proposed network is analyzed with both simulations and analytical expressions.

Another beamforming and \acrshort{mimo} enabled function computation method is given in~\cite{Wen2019}. The study proposes a receiver beamforming model design that reduces channel dimensions and equalizes channel covariances and small scale fading components. Additionally, the method uses a feedback scheme to accurately provide massive \acrshort{csi} knowledge to the network. The proposed approach mainly aims to reduce the computation error resulting from the discrepancy of the received function output. The simulations demonstrate the positive effect of the proposed model on error reduction.

A training model for faster acquisition of the \acrshort{csi} is proposed in~\cite{Ang2019a} and~\cite{Ang2019}. The method is based on the effective \acrshort{csi} definition that is obtained with the simultaneous pilot transmission of the nodes. The method requires iterative broadcasts from \acrshort{fc} to estimate the effective \acrshort{csi} and one last simultaneous pilot transmission yields the \acrshort{csi} vector at the \acrshort{fc} while the traditional \acrshort{afc} methods individually train each \acrshort{csi}. The study analytically obtains the computation complexity of both traditional and proposed methods. Also, an error improvement method is proposed to compensate for the estimation error of the novel approach since the proposed model causes larger estimation errors.  

Function computation is integrated with wireless power transfer in~\cite{Li2019a} for \acrshort{iot} networks. The framework aims to minimize the computation error that is seen at the aggregated data by jointly optimizing power transfer and function computation tasks. The framework is designed for networks that consist of \acrshort{mimo} beamforming capable nodes. The joint optimization problem is divided into two sections; wireless power control optimization and function computation optimization. As applied earlier in~\cite{Chen2018a}, the semidefinite relaxation method is implemented to solve the function computation section of the optimization problem while the wireless power control section is solved in closed form. The study also states that the integration of the wireless power transfer into the \acrshort{afc} network brings an additional design dimension which can increase the computation accuracy. A combination of function computation and wireless power transfer is also implemented for high-mobility sensing in smart cities where \acrshort{uav}s are used to collect sensor readings~\cite{Li2019}.

Power control problem of the \acrshort{afc} systems are considered in~\cite{Cao2019a}. Poor distribution of transmit powers in an \acrshort{afc} network can cause high computation errors as a result of channel distortion. The proposed method aims to find the optimum transmit power level design by solving the optimization problem that minimizes the computation error at the \acrshort{fc}. The definition of the problem involves the optimization of both transmit powers and the denoising factor of the \acrshort{fc}. Moreover, the problem is solved for additional scenarios such that only one transmitter has power constraints instead of all devices. Lastly, the simulations reveal that the proposed power control scheme has a notable mitigating effect on the computation error of the \acrshort{afc} network.


The problem of energy consumption in function computation networks is considered in~\cite{Basaran2020}. The method mainly benefits from the spatial correlations between the sensor readings in order to reduce energy consumption. For this purpose, a minimum mean square error (\acrshort{mmse}) estimator is designed to obtain estimations with less number of samples. The proposed estimator requires significantly less energy consumption to operate, hence nearly doubles the lifetime of the network. In addition to the improved energy efficiency, the estimator also yields better \acrshort{mse} performance compared to the traditional methods as illustrated via simulations.

Wirtinger flow is an algorithm that is usually used in the solution of non-convex optimization problems such as phase retrieval from the received signal magnitudes. Dong \textit{et al.} use the Wirtinger flow method to solve the non-convex optimization problem of the function computation without the knowledge of \acrshort{csi} in~\cite{Dong2020}. The proposed algorithm only requires data samples and randomly initialized Wirtinger flow iterations, i.e. pilot transmission for \acrshort{csi} acquisition is not needed. As a result, the study can reduce the latency of the function computation applications by removing the dependency on the pilot transmission process of the \acrshort{csi} estimation. Also, the study reveals that the estimation error of the Wirtinger flow-based function computation model is sufficiently small.  

In~\cite{Farajzadeh2020}, the power alignment problem of the function computation networks is considered. The power alignment is usually performed with \textit{pre}-processing or pre-coding in traditional function computation networks. In~\cite{Farajzadeh2020}, this problem is addressed with a backscatter framework instead of pre-coding. The proposed method is designed for mass density sensor networks where multiple \acrshort{uav}s are responsible for the collection of sensor measurements. The model consists of two phases; channel gain acquisition and data aggregation. In the first phase, the \acrshort{uav}s are the power emitters and the sensor nodes act as the backscatter object in which the nodes backscatter the ambient signal that comes from the \acrshort{uav}. As a result, the \acrshort{uav} collects the sum channel gain. In the second phase, the \acrshort{uav}s are the readers that receive the aggregated sensor data. The results show that using the sum channel gain for the aggregation of the sensor readings can improve \acrshort{mse} up to 10 dB.

The function computation capability of the \acrshort{afc} is compared with the separate communication schemes in~\cite{Chen2019}. The achievable function rates are derived for the two cases and it is shown and verified via simulations that the computation over the channel (\acrshort{afc}) is not always the optimum scenario for the function computation.

\subsubsection{Test-bed Implementations}

The feasibility of the \acrshort{afc} is also investigated with testbed implementations in the literature. In~\cite{Kortke2014a}, the authors deploy the model of~\cite{Goldenbaum2013b} with \acrshort{sdr}s. A network with 11 nodes and an \acrshort{fc} is implemented to compute the arithmetic and geometric mean of the sensor readings of the nodes. In another implementation study~\cite{Altun2018}, the summation of sensor readings over the channel is tested via \acrshort{sdr} modules. The network that includes three transmitters and an \acrshort{fc} is used to analyze the effect of distance and signal power level. Also, the error performance of the \acrshort{afc} network is compared with the TDMA scheme via computer simulations. The TDMA scheme requires communication slots for each node while the \acrshort{afc} completes the transmission in one slot. As a result, the simulation results of \acrshort{afc} displays less error since TDMA introduces additional thermal noise to the system with each communication. 

In~\cite{Jakimovski2011}, a simultaneous transmission method is proposed and implemented with \acrshort{sdr} modules. The method is based on the hamming distance of the superimposed received signals that are affected by each transmit vector. The calculation of corrupted goods in a pallet via temperature readings is suggested as an application example. The testbed implementation demonstrates the feasibility of the method and the success of computations over the channel. Another implementation study is given by Sigg \textit{et al.} in~\cite{Sigg2012}. They suggested a computation model that carries information in the mean value of the Poisson distribution. The nodes of the system transmit burst sequences that are Poisson distributed and the density of the bursts in a time interval represents the mean of the distribution. The model then extracts the mean value of the superimposed signal at the receiver. The proposed model is implemented with 15 sensor nodes and a receiver that is driven by microcontrollers. The realized scenario successfully recovered the average temperature of the simultaneously transmitted sensor readings.

AirShare method is presented in~\cite{Abari2015} to improve the resistance of the distributed wireless applications against carrier frequency offset (\acrshort{cfo}). The method uses a broadcast clock signal as a reference to other nodes that aim to transmit simultaneously. The paper firstly investigates the feasibility of the AirShare method and then implements the network via \acrshort{sdr} modules.

Superposition of signals is widely considered in the literature for function computation purposes. In these studies, exploiting simultaneous transmission is proven to reduce the computational burden of the receiver since the computations are performed over the air. The major drawback of function computation is the perfect CSI requirement which is investigated in the literature. Also, various proof of concept studies which utilize software defined radios exist in the literature. In conclusion, AFC and AirComp are promising methods that can bring scalability to the computation requirement of a communication system. We expect that emerging low-energy dense IoT networks will be the main target of these methods to solve energy and bandwidth problems.

%% file: federatedlearning.tex
\section{Federated Learning}

An emerging application area of the \acrshort{afc} is the federated learning algorithms which enable the computation of learning data over the air. In this section, the federated learning studies that are based on the simultaneous transmission of signals are examined.

Machine learning (\acrshort{ml}) is the rising technology of the last decade as a result of the increasing computational capabilities of the electronic devices (machines) (comprehensive information on the main contents of this section can be found in~\cite{Zhang2019}). \acrshort{ml} is based on the usage of massive data or data sets to let those machines make the classification or prediction operations. Moreover, many aspects of a communication system such as modulation, demodulation, channel estimation etc. can be considered as a classification or prediction problem~\cite{Gunduz2019}. This observation brought out the relationship between the \acrshort{ml} and wireless communication. Although traditional communication systems use model-based solutions for their problems, \acrshort{ml} based data-driven techniques have already started to show promising results in the wireless communication area~\cite{Simeone2018}. In the following years, this relationship proved to be two ways such that the communication networks can also be beneficial in the learning area. 

The data-driven nature of the \acrshort{ml} requires the collection of massive data in centralized points before the learning process. However, data sources of today's technology are often at the wireless edges. As a result, collecting massive data from wireless devices to a center can cost a high amount of energy and bandwidth~\cite{Amiri2019}. Collaborative machine learning or federated learning is a solution to the data collection problem which proposes the process of the data at the edge users or distributed centers instead of a local center~\cite{Tran2019, Yang2019}. The superposition property of the wireless channel can further relieve some of the costs in the federated learning schemes by performing computations over the wireless channel. A list of federated learning studies can be found in Table~\ref{tab:federatedlearning}.

\begin{table*}[t]  \caption{An overview of federated learning studies.}
  \centering
  \begin{tabular}{ m{4.5cm}|p{1.3cm}|p{8cm}|p{1.6cm}  }
  Author  &  Year  &  Contribution & Performance metric  \\

  \hline \hline

Tran \textit{et al.}~\cite{Tran2019} & \multirow{12}{2cm}{2019}  & The method provides a duality between learning time and energy efficiency.   & Time vs energy cost \\ \cline{1-1} \cline{3-4} 

Yang \textit{et al.}~\cite{Yang2019} &   & Overviews the existing federated learning studies and promotes its data aggregation aspect. & Classification \\ \cline{1-1} \cline{3-4} 

Amiri and Gündüz~\cite{Amiri2019a} &    & Improve the error performance of~\cite{Amiri2019} by compressing the gradient estimate. &  \multirow{10}{2cm}{Accuracy} \\ \cline{1-1} \cline{3-3} 

Amiri \textit{et al.}~\cite{Amiri2019b} &    & Removes the requirement on \acrshort{csi} for the distributed \acrshort{ml} method. &  \\ \cline{1-1} \cline{3-3}

Zhu \textit{et al.}~\cite{Zhu2020} &    & The method is for broadband communications and reduces latency as well as promoting a trade-off between communication and learning performance. & \\ \cline{1-1} \cline{3-3}

Yang \textit{et al.}~\cite{Yang2020} &   & Reduces the convergence rate of the learning algorithm by considering device selection and beamforming. &  \\  \cline{1-3} 

Amiri and Gündüz~\cite{Amiri2019, Amiri2019c, MohammadiAmiri2020} & 2019, 2020   & Analog and digital \acrshort{dsgd} reduce the learning time in bandwidth and energy limited networks. &   \\ 

  \end{tabular}

  \label{tab:federatedlearning}
\end{table*}

\begin{figure}[t]\center
  \includegraphics[width=1\linewidth]{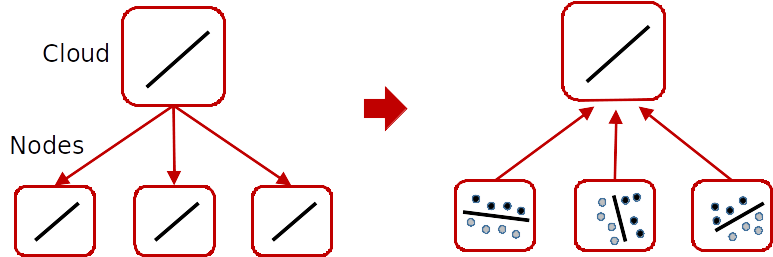}
  \caption{An example of the federated learning scheme. Firstly, cloud distributes the model parameters to nodes. Then each node computes its model update with its own data and transfers the local update to the cloud.}
  \label{fig:feadlearn}
\end{figure}

The usage of simultaneous transmission in federated learning is illustrated in Fig. \ref{fig:feadlearn}. In the first step, cloud broadcasts the network model to nodes (an SVM line is given in the figure.) In the next step, each node computes its local update with its local data. In traditional federated learning schemes, each node transfers its local update to the cloud with pairwise communication. Since the cloud is only interested with the average of the local updates in order to compute the global update, simultaneous transmission can directly give the average over the air.

One fundamental example of this paradigm is given by Amiri and Gündüz~\cite{Amiri2019, Amiri2019c, MohammadiAmiri2020}. The learning algorithm is based on the minimization of a loss function which is solved with the distributed stochastic gradient descent (\acrshort{dsgd}) method. In this method, the learning parameters are updated with multiple iterations. Different from the traditional collaborative \acrshort{ml} that uses \acrshort{dsgd}, this study exploits the wireless channel for the parameter update. In other words, the function that updates the learning parameters are calculated over the air as a result of the superposition property. Two models, digital and analog, are proposed to reduce the required number of iterations to reach an accuracy level. Simulations show that the analog model requires fewer iterations, hence saves both energy and bandwidth of the system. 

In~\cite{Amiri2019a}, Amiri and Gündüz extend their previous model to fading channels. The authors propose the \textit{compressed worker-wise scheduled analog \acrshort{dsgd}} model in which the edge users accumulate the error from previous iterations and reduce the dimension of their transmit vector. The proposed model is also compared with the model of another study that does not reduce the transmit vector (without scheduling) and the results show that the \textit{compressed worker-wise scheduled analog \acrshort{dsgd}} model increases accuracy. In~\cite{Amiri2019b}, the authors remove the CSI knowledge assumption on the transmitters. Instead, the receiver is equipped with multiple antennas. The results show that the increasing number of antennas alleviate the destructiveness of the wireless channel such that the infinite antennas result in totally removed fading and noise. 

Zhu \textit{et al.} take the same approach on federated learning in ~\cite{Zhu2020}; error function is minimized with each iteration and the iterations are calculated over the air. In addition to other studies, tree performance metrics are defined to analyze the network performance and the relations between them are obtained in closed forms. Also, the network is implemented to compare the proposed model with \acrshort{ofdm}. The results confirm the relationships between the defined metrics and show the low latency contribution of the proposed method.

Yang \textit{et al.} consider the same model which is based on the over-the-air computation in~\cite{Yang2020}. The study includes user selection as used in~\cite{Amiri2019a} which schedules the users according to their channel state. As an addition,~\cite{Yang2020} considers the usage of beamforming to improve learning performance. Also, they define and solve the optimization problems that govern the performance of the device selection model. Computation of the input data of a decentralized network is also considered for spectrum sensing algorithms. Different from the federated learning methods, the channel is matched to compute the sensing algorithm functions rather than the learning algorithm functions. In the following section, the studies that benefit from the superposition of signals for the purpose of spectrum sensing are investigated.

Federated learning is the most promising application of simultaneous transmission. By integrating simultaneous transmission (e.g. AFC, AirComp) into federated learning, communication and computation burden can be reduced. Also, simultaneous transmission can also provide  a trade-off between system parameters. Depending on the system parameters (e.g. CSI knowledge level, existence of beamforming, communication band) a proper method can be used from Table \ref{tab:federatedlearning} to improve the federated learning network. Lastly, the major drawback of the federated learning studies comes from the channel estimation error and the literature lacks testbed implementations.

%% file: spectrumsensing.tex
\section{Spectrum Sensing}


Supporting a high amount of users is one of the most important challenges of wireless communication since the frequency spectrum is limited. However, the distribution of the spectrum to the users is usually more challenging than the physical scarcity of the channel itself~\cite{Savage2005}. The cognitive radio is a delicate approach that aims to handle the spectrum access problem efficiently. Cognitive radio is designed to be aware of its surroundings with \acrshort{sdr} capabilities, e.g. spectrum analysis or \acrshort{csi} estimation. The main objective of cognitive radio is to find the idle channels, that is dedicated to primary users, and utilize them to secondary users. The spectrum sensing problem is the first aspect of this objective and attracted distributed network solutions in the literature~\cite{Akyildiz2011}. Some of the simultaneous transmission based spectrum sensing studies are listed in Table~\ref{tab:spectrumsensing} with their contribution and performance metrics.

\begin{table*}[t]  \caption{An overview of spectrum sensing studies.}
  \centering
  \begin{tabular}{ m{2cm}|p{1cm}|p{8cm}|p{2.6cm}  }
  Author  &  Year  &  Contribution & Performance metric  \\

  \hline \hline

Zheng \textit{et al.}~\cite{Zheng2015} & 2015 & The method manages to transfer the sensing data in one time slot, analyze its detection and throughput performance.  & Approximation error, throughput  \\ \hline

Zheng \textit{et al.}~\cite{Zheng2017} & 2017  & Define an energy optimization problem and find the expression of optimal threshold.   & Energy efficiency, sensing time \\ \hline

Chen \textit{et al.}~\cite{Chen2018b} & 2018 & Considers the effect of \acrshort{cfo} in the sensing algorithm. & Signal to aliasing and noise ratio \\

  \end{tabular}

  \label{tab:spectrumsensing}
\end{table*}

The spectrum sensing is a particular case of the detection problem (Section~\ref{sec:detest}) and involves certain network dynamics. In a distributed cognitive radio network, multiple secondary users sense the spectrum and report back to an \acrshort{fc}. A cooperative method that utilizes the simultaneous transmission for this problem is proposed in~\cite{Zheng2015}. Instead of individual sensing, the method takes help from the decentralized nodes to detect the free frequency bands. Spectrum sensing mechanism is based on observed energy from the frequency channels. This energy information is carried simultaneously from the distributed nodes to the \acrshort{fc} as encoded to the signal energies and the final decision is made at the \acrshort{fc}. The detection probability and the false alarm probability of the proposed method are derived and used in order to optimize the detection performance. Also, the proposed scheme is compared with the traditional spectrum sensing methods with respect to the approximation error and network throughput via simulations. The authors later consider the energy efficiency of the spectrum sensing problem in~\cite{Zheng2017}. For this reason, an optimization problem is derived that considers the sensing time, the detection threshold and the symbol sequence length. The problem is simplified, solved and the results are verified with simulations.

In~\cite{Chen2018b}, the proposed cognitive radio network focuses on the wideband spectrum. The objective of the method is to find the occupied wideband channels with low latency and high accuracy. The method utilizes a distributed network model to improve the sensing accuracy and benefits from the superposition of the signals to reduce the delay time. Specifically, the discrete Fourier transform (\acrshort{dft}) of the data is computed over the nodes and the air. The network is examined for the synchronization errors (e.g. synchronization phase offset) that occur between the nodes. Furthermore, a robust estimation and equalization technique is proposed to mitigate the effects of the imperfect synchronization. The performance of the proposed method is analyzed theoretically and verified with both simulation and \acrshort{sdr} based implementation. The computation of functions over the air is also extended for detection and estimation algorithms to reduce the latency and improve the error performance. In these studies, the channel is adopted to collect the data from distributed sources and process them simultaneously over the air for the detector or estimator.

%% file: detectionestimation.tex
\section{Detection and Estimation} \label{sec:detest}

Detection and estimation are essential parts of many engineering applications and their performance profoundly affects the performance of the system. While the detection and estimation theory is concerned with the optimal design of the decision mechanisms, the acquisition of the samples is another important parameter for the whole system. In general, larger sample sizes produce better decision performance, however, it requires more energy consumption and long computation times. For this reason, the response of the decision performance to the sample size is an important aspect to consider. The \textit{error exponent} is a measure that indicates how fast the error changes with the increasing sample size and is often used in the analysis of asymptotic behavior of the sample size. 
This aspect is particularly important in the \acrshort{iot} applications that can involve massive sensor networks in which the observations from multiple sensors together create the sample set. For this reason, distributed detection or estimation strategies are developed to improve the performance of this process. In this section, we present the distributed detection and estimation studies that exploit simultaneous transmission.

\subsection{Detection}


Several detection studies based on simultaneous transmission are presented in Table~\ref{tab:det}. The distributed detection problem of wireless networks is considered in~\cite{Liu2007}. The network consists of spatially distributed sensors that aim to transfer the statistics of the sensor measurements to the \acrshort{fc}. The method exploits the superposition property of the \acrshort{wmac} for this purpose by simultaneously transmitting their statistics. The study examines the detection performance of the proposed method from the perspective of the number of measurements and power consumption. Two sensor types are considered for the review; intelligent and dumb. The intelligent sensors are aware of the source statistics and transfer the log-likelihood ratio (\acrshort{llr}) to the \acrshort{fc}. It is shown that the \acrshort{llr} based detection is asymptotically optimal, i.e. it can reach the performance of the centralized detection. The dumb sensors are unaware of the source statistics and transfer the histogram of their measurements to the \acrshort{fc}, however, source statistics are required at the \acrshort{fc}. The results show that the histogram transfer is also asymptotically optimal. The results are also verified with simulations that show the detection error as a function of the number of sensors.

\begin{table*}[t]  \caption{An overview of detection studies.}
  \centering
  \begin{tabular}{ m{3cm}|p{1cm}|p{9cm}|p{2.3cm}  }
  Author  &  Year  &  Contribution & Performance metric  \\

  \hline \hline

Liu and Sayeed~\cite{Liu2007} & 2007 & A low-complexity distributed detection mechanism (based on type function) is proposed and investigated for the error exponent. & Error probability  \\ \hline

Mergen \textit{et al.}~\cite{Mergen2007} & 2007  & Provides an asymptotically optimal detector and analyzes its error exponent performance.   &  \multirow{6}{2cm}{Error exponent}  \\ \cline{1-3}

Li and Dai~\cite{Li2007} & 2007 & Proposes a bandwidth and delay efficient method and compares with the separation based methods. &  \\ \cline{1-3}

Li \textit{et al.}~\cite{Li2012} & 2011  & Analyzes and compares the \acrshort{mdf} and the \acrshort{maf} methods for error exponents under power constraints. & \\ \cline{1-3}

Banavar \textit{et al.}~\cite{Banavar2012} & 2012  & Includes multiple antenna receivers and investigates different \acrshort{csi} scenarios.  &  \\ \hline

Ralinovski \textit{et al.}~\cite{Ralinovski2016} & 2016  & The anomaly detection method reduces the communication costs (energy, bandwidth) and investigate relation between accuracy and communication costs.  & Reliability, energy consumption \\ \hline

Raceala-Motoc \textit{et al.}~\cite{Raceala-Motoc2019} & 2018  & Gives the upper-bounds for the probability of mislabeling for~\cite{Ralinovski2016}.   & Probability of mislabeling \\

  \end{tabular}

  \label{tab:det}
\end{table*}

Sensor measurements of a wireless sensor network are used for target detection in~\cite{Mergen2007}. The sensor measurements are transferred to the test center via type-based multiple access where the same frequency and time resources are allocated to all users. In \acrshort{tbma}, the receiver only accesses to the histogram of the transmitted data (we refer the reader to Section~\ref{sec: multipleaccess} for detailed information) and the detection mechanism is based on the histogram of the observations. The study focuses on the performance analyses of \acrshort{tbma} based detection schemes (includes a large deviations approach). An asymptotically optimal detection mechanism is proposed and its detection error exponents are derived. Channels with \acrshort{iid} and non-\acrshort{iid} gains are considered for the evaluations. Also, the error probabilities of \acrshort{tbma} scheme is compared with \acrshort{tdma} via simulations.

In~\cite{Li2007}, the existence of a signal is checked with a binary hypothesis test where the observation samples are obtained from a decentralized sensor network. The sensor readings are collected in the \acrshort{fc} via two-channel models; each sensor has its own dedicated channel or a single channel is dedicated to all sensors. The second model that dedicates a single channel to the sensors attracts our interest since it allows the superposition of the signals. The study focuses on the detection performance of these two-channel models and their comparison with the centralized detection model. In the analyses, two Gaussian noise cases (correlated and independent) and two power constraint cases (average and total) are considered. Bayesian error exponents of the detection probabilities are derived for these scenarios (includes large deviations approach). Several results are presented in the study and verified via simulations that examine the effect of the number of sensors on the detection error probability and error exponent. The result for the average power constraint case is that the superimposed signals with the correlated Gaussian noise can reach the error performance of the centralized detection scenario. However, dedicating channels to each sensor always leads to error performance reductions. For the total power constraint case, an increasing number of sensors exponentially reduces the error exponent for the superimposed signals.  

Two superposition based schemes are proposed in~\cite{Li2012} for the distributed signal detection via binary hypothesis testing. The modified detect-and-forward (\acrshort{mdf}) scheme considers the signal detection at each distributed node and transfers the test results to the \acrshort{fc}. The second scheme, modified amplify-and-forward \acrshort{maf}, transfers the observations to the \acrshort{fc} before the hypothesis testing. The study examines the detection performance of these schemes under individual power constraint and total power constraint. The results show that the \acrshort{maf} is asymptotically (i.e. infinite number of sensors) optimal under individual power constraint, however \acrshort{mdf} is not optimal. 

In~\cite{Banavar2012}, multiple antennas at the \acrshort{fc} is considered for the distributed detection problem. The upper and lower bounds of the error exponents are derived for the scenarios: \acrshort{awgn} channels, Rayleigh channels, full \acrshort{csi} enabled, phase-only \acrshort{csi} enabled. It is shown that equipping the \acrshort{fc} with multiple antennas presents a gain of $\pi/8$ for the scenario of Rayleigh channel and full \acrshort{csi}. Four algorithms are proposed for the design of the sensors' power allocation. Lastly, the error exponents of these algorithms are compared with the derived bounds via simulations.

An anomaly detection algorithm is proposed in~\cite{Ralinovski2016} which is based on the \acrshort{afc} given in~\cite{Goldenbaum2013b}. A supervised learning algorithm is used for the hypothesis testing and the classifier of the algorithm is generated at the \acrshort{fc} with the sensor readings. Since the \acrshort{fc} is only interested in a function of the sensor readings, \acrshort{afc} is proposed for the classifier generation over the channel. The main contribution of the study is its energy efficiency and it is shown that the proposed scheme can significantly reduce the consumed energy compared to the \acrshort{tdma} scheme. Later, in~\cite{Raceala-Motoc2019}, the study on anomaly detection is extended to include the upper bounds on the probability of mislabeling.

\subsection{Estimation}


A list of the estimation studies that benefit from the superposition of signals can be found in Table~\ref{tab:est}. An estimation method that is based on the \acrshort{tbma} scheme is proposed in~\cite{Mergen2006} . The method benefits from the superposition of the sensor readings over the channel to gain from the bandwidth and the delay time. The study aims to design the optimum estimation process that includes the transfer of the samples to the \acrshort{fc} and the estimator. The study derives the Cramer-Rao bound of the estimation problem and designs the estimation process as the combination of \acrshort{tbma} and an \acrshort{ml} detector. The results show that the \acrshort{tbma} based estimation is asymptotically optimal if the channel gains of all sensors are the same. Additionally, the proposed model is analyzed for the fading channels and compared with the \acrshort{tdma} based schemes via simulations.

\begin{table*}[t]  \caption{An overview of estimation studies.}
  \centering
  \begin{tabular}{ m{3cm}|p{1cm}|p{9cm}|p{2.3cm}  }
  Author  &  Year  &  Contribution & Performance metric  \\

  \hline \hline

Mergen and Tong~\cite{Mergen2006} & 2006 & Analyzes the asymptotic behaviour of \acrshort{tbma} based estimation.  &  \multirow{3}{2cm}{\acrshort{mse}}  \\ \cline{1-3}

Xiao \textit{et al.}~\cite{Xiao2006} & 2008  & Propose a bandwidth and power efficient method and defines and solves an optimization problem for power scheduling.   & \acrshort{mse} \\ \hline

Bajwa \textit{et al.}~\cite{Bajwa2007} & 2007 & Proposes a energy efficient joint source-channel based method and defines the relationship between its power consumption, error rate and latency. & Power, distortion, scaling exponents \\ \hline

Wang and Yang~\cite{Wang2010} & 2010  & The method removes the \acrshort{csi} requirement by using more bandwidth. & \acrshort{mse}, bandwidth \\ \hline

Banavar \textit{et al.}~\cite{Spanias2010} & 2010  & Considers and investigates different fading models and their effect on the performance. & Asymptotic variance \\

  \end{tabular}

  \label{tab:est}
\end{table*}

In~\cite{Xiao2006}, power and bandwidth limited networks are considered for the distributed estimation of unknown signals. The authors especially propose a power scheduling model and investigate the cases where the observations are scalar and vectorial. The optimization problem for power scheduling is shown to be convex and solved for the scalar observations. Additionally, it is shown via simulations that the proposed scheduling generates better \acrshort{mse} performance than the uniform scheduling. The optimal scenario to estimate the vectorial observations is investigated for both noiseless and noisy channel models. The results for the noiseless channel model are given in closed form; however, the solution of the optimization problem for the noisy channel required semidefinite relaxation methods.

A joint source-channel communication structure is considered for decentralized wireless sensor networks in~\cite{Bajwa2007}. The objective of the study is to build a communication infrastructure that reduces bandwidth and power consumption. For this purpose, optimization of the acquisition, communication and processing of the measured data is considered collectively. The relationships between the power consumption, error rate and the latency of the network are derived for the increasing number of sensors and verified with simulations. It is shown that the efficient and healthy estimation of the unknown signals is possible for a large number of sensors if prior knowledge is allowed in the proposed approach. Moreover, the method still produces healthy estimation results when partial or no prior knowledge is given if the power and latency constraints are increased sublinearly as a function of the number of sensors.

A \acrshort{tbma} based estimation method is proposed in~\cite{Wang2010} for distributed \acrshort{wsn}s. In traditional \acrshort{tbma} schemes; the fading of the wireless channel notably degrades the communication performance and providing \acrshort{csi} to each node via pilot transmission or feedback channel brings excessive burden for a large number of sensors. The proposed method aims to provide robustness against these problems by partitioning additional bandwidth to the network. The main idea behind the model is to use multiple orthogonal waveforms for each message (type) where the traditional \acrshort{tbma} uses only one. The authors also designed a maximum likelihood estimator for the system and derived the Cramer-Rao lower bound expression. The simulations compare the \acrshort{mse} performance of the method with traditional \acrshort{tbma} and illustrate the relationship between the bandwidth, \acrshort{snr} and the number of nodes.

Fading characteristics of the channel and the amount of allowed \acrshort{csi} at the network is highly important for the performance of the distributed sensor networks that exploit the superposition of the signals. In~\cite{Spanias2010}, various fading models and different \acrshort{csi} amounts are tested for distributed estimation networks. The variance expression for the network's estimate is derived for the perfect and partial \acrshort{csi} cases. Also, different channel characteristics are considered and the corresponding variance expressions are derived for a large number of sensors. The authors consider the impact of the errors that occur at the feedback process in their evaluations. The convergence rates are calculated for several scenarios. It is shown that the asymptotic results can be nearly reached with a practical number of sensor nodes. The results are verified with computer simulations. Simultaneous transmission is an effective tool for the consensus algorithms that reduce the convergence time of the consensus process. In the following section, the studies that exploit the superposition property of the \acrshort{wmac} are presented.

%% file: gossipconsensus.tex
\section{Gossip and Consensus}

The physical layer is generally exploited for many-to-one networks where an \acrshort{fc} requires the message of multiple users ($K \times 1$). However, in consensus problems, all users in the network aim to agree upon a common value, e.g. average, which carries a much higher communication burden. Finding an efficient communication structure for the consensus problem that provides fast and reliable convergence, while requiring low-energy, is a challenging task~\cite{Olfati-Saber2007}. Gossiping is a distributed form of consensus that the nodes locally communicate with their neighbors instead of following a network-wide protocol. It has been shown that simultaneous transmission presents opportunities for the improvement of gossip and consensus algorithms. A list of existing consensus studies is given in Table~\ref{tab:consensus}. In this section, we examine the consensus studies that exploit the wireless channel to provide energy and time efficiency. 

\begin{table*}[t]  \caption{An overview of consensus studies.}
  \centering
  \begin{tabular}{ m{3.3cm}|p{1cm}|p{8.7cm}|p{2.3cm}  }
  Author  &  Year  &  Contribution & Performance metric  \\

  \hline \hline

Kirti \textit{et al.}~\cite{Kirti2007} & 2007 & Provides a scalable consensus algorithm that the convergence time is independent of the network size. & \multirow{5}{2cm}{\acrshort{mse}}  \\ \cline{1-3}

Goldenbaum \textit{et al.}~\cite{Goldenbaum2012} & 2012  & The consensus of multiple nodes with respect to nomographic functions is established and analyzed.  &  \\ \cline{1-3}

Steffens and Pesavento~\cite{Steffens2012} & 2012  & Provides a low-complexity and scalable consensus algorithm.  &  \\ \hline

Nazer \textit{et al.}~\cite{Nazer2009} & 2009  & Faster than the pairwise gossip and provides energy efficiency exponential to the network size.  & \multirow{3}{2cm}{Number of gossip rounds}  \\ \cline{1-3} 

Nazer \textit{et al.}~\cite{Nazer2011a} & 2011 & Provides time and energy efficiency polynomial to the network size. & \\ \hline

Nokleby \textit{et al.}~\cite{Nokleby2011} & 2011  & Based on full duplex communication and provides scalability to the network size. & Averaging time \\ \hline

Molinari \textit{et al.}~\cite{Molinari2018} & 2018  & Robust to fading and provides fast convergence. & Convergence time \\ \hline

Agrawal \textit{et al.}~\cite{Agrawal2019} & 2019  & Provides robustness to noise and low \acrshort{snr} for max-consensus networks. & Error rate, number of iterations \\

  \end{tabular}
  \label{tab:consensus}
\end{table*}

The average consensus problem of wireless networks is considered in~\cite{Kirti2007}. The method exploits the physical layer with the simultaneous transmission in order to update the consensus algorithm. The results show that each node in the proposed network can obtain the average of the data under a sufficient \acrshort{mse} level. Since the channel is accessed by the nodes simultaneously, the method presents better convergence performance compared to the conventional consensus algorithms as the number of nodes increase.

In~\cite{Nazer2009}, Nazer \textit{et al.} use the computation coding in a gossip algorithm that averages the sensor readings. The method is based on the simultaneous transmission of the nodes by exploiting the superposition property of the physical layer. The network is arbitrarily divided into local neighborhoods such that the average consensus is initially provided for each of neighborhood, later the global consensus is achieved for the network. The results reveal that the energy efficiency of the network increases exponentially and the time efficiency increases polynomial as the number of nodes increase. In~\cite{Nazer2011a}, the previous studies are extended and compared with the nearest neighborhood gossip algorithms via simulations. Also, it is shown that the proposed method converges in $O(n^2/m^2)$ rounds while the nearest neighborhood algorithm converges in $O(n^2)$ rounds where $n$ and $m$ are the network and local neighborhood sizes respectively. 

The averaging gossip problem of a wireless network is extended to nomographic functions in~\cite{Goldenbaum2012}. The main idea behind the study is the analog computation of the nomographic consensus functions over the wireless channel with the concurrent transmission. The proposed model is cluster/neighborhood-based as in~\cite{Nazer2009} and~\cite{Nazer2011a} such that the consensus is initially provided for local clusters rather than the whole network. Different from~\cite{Nazer2009} and~\cite{Nazer2011a}, the global consensus is obtained with the existence of the common nodes which connect the clusters with each other. Two algorithms, deterministic and randomized, are proposed and it is shown that both of them significantly increase the convergence rate.

Another gossip algorithm that exploits the physical layer is proposed in~\cite{Nokleby2011}. The novelty of the study lies in the full-duplex communication model of the network which allows simultaneous transmission and reception of signals with special self-interference cancellation techniques. Each node in the network broadcasts to its neighbors and receives from them simultaneously and only requires synchronization. The theoretical results show that the proposed gossip algorithm has a three times faster convergence rate than the randomized gossip algorithms. Also, the study includes computer simulations supporting the presented theoretical results.

In~\cite{Steffens2012}, the average consensus of a wireless network is considered. Although many of the simultaneous transmission based consensus algorithms are neighborhood/cluster-based (as in~\cite{Nazer2009,Nazer2011a,Goldenbaum2012}), this study arbitrarily divides the network into two subgroups rather than many. In each iteration, all the nodes of a subgroup simultaneously broadcast their data while the nodes in the other subgroup receive the superimposed signal. After an iteration, the subgroups change roles such that the receiver subgroup becomes the transmitting subgroup and vice versa. The nodes in the transmitting subgroup create their new data as a weighted sum of the received signal and the original data. The proposed method offers energy efficiency and significant convergence performance as the simulations illustrate the comparison of the proposed method with the randomized and the broadcast gossip algorithms.

The superposition of the transmitted signals is also considered in~\cite{Molinari2018}. The study uses similar grounds that are used in~\cite{Goldenbaum2012}.~\cite{Molinari2018} also includes unknown channel coefficients in the design of the consensus method. Time-variant and time-invariant cases are considered for the wireless channel model. A tuning parameter (a stubbornness index) is proposed to control the convergence rate. The results show that the smaller values of the tuning parameter (stubborn systems) reduce the effect of the time-variant channel coefficients and increase the convergence time. However, the tuning parameter only affects the convergence rate in the time-invariant systems. The study also verifies the theoretical results via simulations. A max consensus scheme (\textit{ScalableMax}) is proposed in~\cite{Agrawal2019}. The method focuses on the dense networks where a large number of nodes aims to find the maximum function. The main contribution of the proposed method is its scalability. Moreover, an error correction mechanism is added to the system in order to increase the system's resilience to the low \acrshort{snr} region. The \acrshort{cpnf} and \acrshort{afc} also inspired security application in the literature. In the following section, we present the security applications that benefit from simultaneous transmission. 


%% file: security.tex
\section{Security}

Superposition of signals is shown to be beneficial to provide security to wireless networks. Several security studies that use simultaneous transmission are listed in Table~\ref{tab:security}. In this section, we present the security applications in detail.

\begin{table*}[t]  \caption{An overview of security studies.}
  \centering
  \begin{tabular}{ m{3.7cm}|p{1cm}|p{8.7cm}|p{2.4cm}  }
  Author  &  Year  &  Contribution & Performance metric  \\

  \hline \hline

Shashank and Kashyap~\cite{Shashank2013} & 2013 & Achieves strong secrecy with lattice codes and randomized encoding. & \multirow{2}{2cm}{Achievable power-rate pair}  \\ \cline{1-3}  

Vatedka \textit{et al.}~\cite{Vatedka2015} & 2015 & Provides perfect or strong secrecy to two-way relay networks using \acrshort{cpnf}.  &  \\ \hline

Babaheidarian and Salimi~\cite{Babaheidarian2015} & 2015  & Combines lattice alignment and asymmetric \acrshort{cpnf} to improve the secure sum rates. & Achievable sum rate \\ \hline

Richter \textit{et al.}~\cite{Richter2015} & 2015  & Includes fading and provides weak secrecy for multi-way relay networks using \acrshort{cpnf}.  &  \multirow{6}{2cm}{Secrecy capacity}  \\ \cline{1-3} 

Karpuk and Chorti~\cite{Karpuk2016} & 2016  & Derive the secrecy rate upper bounds, investigates the effect of synchronization error and extends the proposed model to \acrshort{mimo} sceheme. &  \\ \cline{1-3} 

Ren \textit{et al.}~\cite{Ren2017} & 2017  & Investigates internet and external eavesdropper scenarios, includes two-hop channel, jammer and improves the secrecy rate. & \\ \hline

Goldenbaum \textit{et al.}~\cite{Goldenbaum2016a} & 2016  & Defines and derives the secrecy computation-capacity and shows that it can be achieved without sacrificing capacity. & \multirow{4}{2cm}{Secrecy computation-rate}  \\ \cline{1-3} 

Goldenbaum \textit{et al.}~\cite{Goldenbaum2017} & 2016  & Secure against the eavesdroppers that has good channel conditions and applicable with low-complexity. & \\ \hline

Babaheidarian \textit{et al.}~\cite{Babaheidarian2017} & 2017  & Considers the scenario of malicious receiver and provides secrecy in this scenario with beamforming and jamming. & Achievable secure rate. \\ \hline

Hynek Sykora~\cite{Hynek2015} & 2015  & Uses game theory to provide secrecy. & Pay-off matrix \\ \hline

Negi and Goel~\cite{Negi2005} & 2016  & Provides secrecy by using artificial noise for networks with multiple antenna or helper nodes. & Secrecy Capacity, outage probability \\ \hline

Altun~\cite{Altun2020} & 2020  & Obtains individual information at the \acrshort{fc} and provides authentication to the transferred messages. & Probability of detection and false alarm  \\

  \end{tabular}
  \label{tab:security}
\end{table*}

The \acrshort{cpnf} method is utilized for the security purposes in~\cite{Shashank2013}. The study considers a two-transmitter network that communicates with a curious relay using \acrshort{cpnf} over the \acrshort{awgn} channel. The security of the method is confirmed by investigating the mutual information between the individual messages and the superimposed signal at the relay. It is shown that the mutual information is significantly small for large block lengths and gives insufficient information to the relay. The authors also derive the achievable rates and the necessary conditions for secure communication. Lastly, the network is generalized for the multi-hop networks.

Another secure communication scheme based on the \acrshort{cpnf} is given in~\cite{Babaheidarian2015}. In the study, an asymmetric \acrshort{cpnf} model (as in~\cite{Ntranos2013}) is considered which assumes asymmetric channel gains towards the eavesdropper and the secure sum rates are derived.

A comprehensive study on secure communication is given in~\cite{Vatedka2015} which is based on the \acrshort{cpnf} over bi-directional relay channels. Two nodes in the proposed model aim to communicate with the help of a relay without leaking information to the relay. Perfect and strong secrecy conditions are defined to evaluate the secrecy performance. It is shown that the proposed lattice coding and the \acrshort{cpnf} provide both strong and perfect secrecy even in a noiseless design. Also, the results reveal that the noise only affects the computation performance. Lastly, the computation rate under the Gaussian noise is derived. Another comprehensive secure communication study is given in~\cite{Richter2015}. In addition to~\cite{Vatedka2015}, this study considers single-input-multiple-output (\acrshort{simo}) multi-user multi-way networks and includes fading to the channel model. The study derives the secrecy region under the weak secrecy condition. The results show that the securely achievable sum-rate is equal to the difference between the computation rate and the \acrshort{mac} capacity. The simulations verify the derived results. Also, the proposed model is compared with the traditional insecure \acrshort{cpnf} and the security schemes of~\cite{Vatedka2015} from the secrecy rate perspective. 

Secure communication of two nodes through a relay node is considered in~\cite{Karpuk2016} with the help of \acrshort{plnc}. The study considers perfect secrecy conditions and defines the upper bounds on the achievable secrecy rate for noiseless channels. Different coding algorithms are designed for scenarios where the nodes are alone or they cooperate. Moreover, the achievable rates are calculated for these scenarios and it is shown that the given algorithms reach close to the calculated upper bounds. Lastly, the study is extended to a scenario, where the nodes are equipped with multiple antennas. 

In~\cite{Goldenbaum2016a}, the secure communication of multiple nodes is aimed in a modulo-$2$ adder network. The objective is to leak no information to the eavesdropper while the legitimate receiver obtains the modulo-$2$ sum of the transmitters. The study defines the secrecy computation communication capacity which is the maximum achievable secure computation rate and it is shown that the secrecy computation communication capacity is the same as the computation capacity under certain conditions. In other words, the proposed scheme can achieve the security constraints without reducing the capacity. 

Secure function computation based on the \acrshort{afc} over a wiretap channel is considered in~\cite{Goldenbaum2017}. The study aims to keep the eavesdropper ignorant and contributes to the complexity such that the method can provide security without the need for additional stochastic encoding. Also, the study assumes no advantage over the eavesdropper e.g. better channel quality.

In~\cite{Ren2017}, a modified \acrshort{cpnf} given in~\cite{Zhu2014a} is exploited for the purpose of secure communication. Two scenarios where the adversary is external and internal (the relay) are considered. In the internal case, the relay is assumed to be curious that eavesdrops the received signal and the destination node is assumed to be cooperative by jamming the relay to prevent information leakage. The random binning and the lattice chain codes are used in the proposed scheme. It is shown that the proposed methods can achieve the secrecy capacity at the high \acrshort{snr} regions. Moreover, the results revealed that the obtained secrecy capacity is close to the channel capacity. Specifically, the security constraints only lower the insecure channel capacity by $1/2$ bits per channel use. Cooperative jamming is also applied in~\cite{Babaheidarian2017}. Secure communication from multiple sources to multiple receivers is aimed with the help of multiple relays. Contrary to~\cite{Ren2017}, the relays are assumed to be trustworthy and cooperative in the way that the relays use beamforming to jam the malicious receivers.

\cite{Hynek2015} considers a scenario such that the relays of a \acrshort{plnc} network can be malicious and intentionally use a deceptive mapping. The authors view this problem as an incomplete information game and derive an equilibrium. In~\cite{Negi2005}, Negi and Goel propose a secure transmission model that uses artificial noise. The proposed model intentionally introduces AN to the wireless channel in order to degrade the eavesdropper's channel. Naturally, the added AN also degrades the legitimate receiver's channel. This problem is solved by separating the AN source from the signal source. The authors propose two methods for the separation; using other antennas (multiple antenna) or other users (helpers). In both methods, AN and signal source transmit simultaneously at the same time and frequency slot. However, AN source chooses the noise vector such that the noise and the channel coefficient of the legitimate receiver cancel each other. In other words, AN lies in the null space of the channel coefficient of the legitimate receiver. The paper makes two critical assumptions. Firstly, transmitters perfectly know the receiver's channel. Secondly, legitimate receiver's and the eavesdropper's channels are uncorrelated, which enables AN to degrade any channel other than the legitimate receiver's. Presented methods are also supported and compared with simulations.  

An authentication method is proposed in~\cite{Altun2020}. The authors consider an uplink communication scenario where multiple users simultaneously transmit their information to a receiver. Contrary to the conventional \acrshort{afc} studies, this method enables the reconstruction of individual data at the receiver. For this purpose, unique \textit{pre} and \textit{post} processing functions that are based on Gaussian primes are proposed. As a result, the method can simultaneously transfer the individual data of multiple users to the receiver. More importantly, the authors argue that the \textit{post}-processing function also enables the receiver to authenticate the legitimate transmitters. The reason behind this argument comes from the uniqueness of the fading coefficients between any two points and the uniqueness of the Gaussian primes. The study firstly proves the feasibility of the proposed approach with numerical results and later investigates the security aspect of the method by simulating the probability of detection and false alarm metrics.